\documentclass[prx,twocolumn,showpacs,superscriptaddress,preprintnumbers,amssymb]{revtex4-2}
\usepackage{graphicx}
\usepackage{latexsym}
\usepackage{amsmath}
\usepackage{mathtools}
\usepackage{amsfonts}
\usepackage{upgreek}
\usepackage{bm}
\usepackage{multirow}
\usepackage{enumitem}
\usepackage{color}
\usepackage[colorlinks, citecolor=blue]{hyperref}
\usepackage{physics}
\usepackage{tcolorbox}
\usepackage{manfnt}
\usepackage{relsize}

\newcommand{\beq}{\begin{equation}}
\newcommand{\eeq}{\end{equation}}
\newcommand{\beqn}{\begin{eqnarray}}
\newcommand{\eeqn}{\end{eqnarray}}

\newcommand{\vphi}{\varphi}

\newcommand{\secref}[1]{Sec.\,\ref{#1}}
\newcommand{\appref}[1]{Appendix.\,\ref{#1}}
\newcommand{\eqnref}[1]{Eq.\,\eqref{#1}}

\newcommand{\figref}[1]{Fig.\,\ref{#1}}

\newcommand{\rAngle}{\rangle \hspace{-2pt} \rangle }
\newcommand{\lAngle}{\langle \hspace{-2pt} \langle }

\usepackage{slashed}

\begin{document}

\title{Information Critical Phases under Decoherence}

\author{Akash Vijay}
\affiliation{Department of Physics and Anthony J. Leggett Institute of Condensed Matter Theory, University of Illinois at Urbana-Champaign, Urbana, Illinois 61801, USA}

\author{Jong Yeon Lee}
\affiliation{Department of Physics and Anthony J. Leggett Institute of Condensed Matter Theory, University of Illinois at Urbana-Champaign, Urbana, Illinois 61801, USA}
\affiliation{Korea Institute for Advanced Study, Seoul 02455, South Korea}

\date{\today}
\begin{abstract}
Quantum critical phases are extended regions of phase space characterized by a diverging correlation length. By analogy, we define an \emph{information critical phase} as an extended region of a mixed state phase diagram where the Markov length, the characteristic length scale governing the decay of the conditional mutual information (CMI), diverges. 

We demonstrate that such a phase arises in decohered $\mathbb{Z}_{N}$ toric codes by assessing both the CMI and the coherent information, the latter quantifying the robustness of the encoded logical qudits. 
For $N>4$, we find that the system hosts an information critical phase intervening between the decodable and non-decodable phases where the coherent information saturates to a fractional value in the thermodynamic limit, indicating that a finite fraction of logical information is still preserved.  We show that the density matrix in this phase can be decomposed into a convex sum of Coulombic pure states, where gapped anyons reorganize into gapless photons. We further consider the ungauged $\mathbb{Z}_{N}$ toric code and interpret its mixed state phase diagram in the language of strong-to-weak spontaneous symmetry breaking. We argue that in the dual model, the information critical phase arises because the spontaneously broken off-diagonal $\mathbb{Z}_{N}$ symmetry gets enhanced to a $U(1)$ symmetry, resulting in a novel superfluid phase whose gapless modes involve coherent excitations of both the system and the environment. Finally, we propose an optimal decoding protocol for the corrupted $\mathbb{Z}_{N}$ toric code and show that its logical error rate saturates to a fractional value in the information critical phase. Our findings identify a gapless analog for mixed-state phases that still acts as a fractional topological quantum memory, thereby extending the conventional paradigm of quantum memory phases. 
\end{abstract}
\maketitle

\tableofcontents

\begin{figure*}[!t]
    \includegraphics[width=0.99\textwidth]{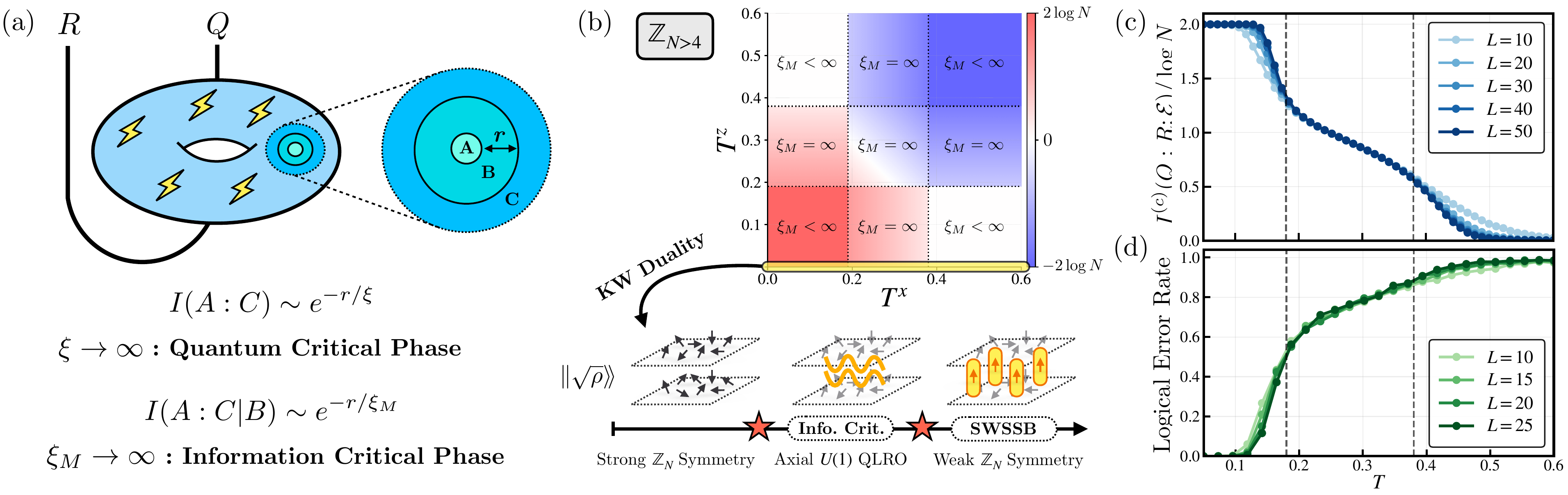}
    \caption{\label{fig:Summary_Figure} \textbf{(a) Information critical phase.} 
    The Markov length $\xi_M$ sets the length scale for the spatial decay of the conditional mutual information $I(A\,{:}\,C|B)$, analogous to how the correlation length $\xi$ controls the decay of the mutual information $I(A\,{:}\,C)$. An \emph{information critical phase} is an extended region of phase space where $\xi_M=\infty$ while $\xi<\infty $.
    \textbf{(b) Phase Diagram.} 
    We chart the mixed-state phase diagram of the $\mathbb{Z}_{N>4}$ toric code as a function of effective temperatures $T^x$ and $T^z$, which parametrize the strengths of $X$ and $Z$ dephasing, respectively. Throughout the diagram $\xi=0$ while $\xi_{M}$ diverges over extended regions. In the ungauged, Kramers–Wannier (KW) dual of the $\mathbb{Z}_N$ toric code, the decodability transition corresponds to a strong-to-weak SSB transition. The dual spin model evolves from a strongly $\mathbb{Z}_N$-symmetric phase into an intermediate phase where the broken axial $\mathbb{Z}_N$ symmetry effectively gets enhanced to a $U(1)$ symmetry. This gives rise to a novel superfluid phase with quasi-long-range order (QLRO). Further increasing the decoherence strength eventually causes full SWSSB of the $\mathbb{Z}_N$ symmetry.
    \textbf{(c) Coherent information.} For $N=8$, we plot the normalized $I^{(c)}$ as a function of $T$ across different system sizes. It approaches its maximal value for $T<0.185$, and remains finite in the information-critical regime $0.185<T<0.38$, indicating \emph{fractionally recoverable} logical information. For $T>0.38$, $I^{(c)}$ decays to zero.
    {\bf (d) Logical error rate.} We implement maximum-likelihood decoding via a Monte Carlo worm algorithm combined with parallel tempering. The resulting decoder exhibits error thresholds that coincide precisely with the critical points identified from the coherent information. Within the information critical phase, the logical error rate saturates to a nonzero fractional value in the thermodynamic limit, providing an operational signature of a fractionally decodable memory phase.
    }
\end{figure*}

\section{Introduction}

Quantum information is inherently fragile in the presence of environmental noise. Thus, the formulation and design of robust quantum error-correcting codes is an indispensable aspect of any viable theory of quantum computation~\cite{Schor_1995,CSS_1996,Steane_CSS_Codes,Laflamme_Perfect_QECC, Gottesman_Intro_QECC}. 
Topological codes, in particular, offer a promising route to realize fault-tolerant quantum memories and scalable quantum processors~\cite{Freedman_Kitaev_Topological_Computation,Kitaev_Topological_Error_Correction,Surface_Codes_Definition, Review_Quantum_Memories,Review_Non_Abelian_Anyons, Color_Codes, Subsystem_Codes, Haah_Codes, Gauge_Color_Codes}. They also provide a physical realization of novel phases of matter with ground state degeneracy and long range entanglement~\cite{Wen_Topological_Order_Rigid_States,String_net_Condensates,Levin_Wen_TEE,Kitaev_Preskil_TEE,Book_XGWen_Many_Body_Physics, Wen_Colloquium}. 
Considerable effort has gone into determining the error thresholds of topological quantum codes~\cite{Dennis_2002_Error_Threshold,Wang_2003_Error_Threshold,Color_Codes_Error_Threshold,Depolarization_Error_Threshold,3D_Color_Code_Threshold}. Early studies operationally defined these thresholds by assessing the success or failure of specific decoding protocols. More recent work reframes the decodability transition as an \emph{intrinsic} mixed-state phase transition, across which information-theoretic quantities exhibit a sharp, non-analytic jump~\cite{Fan_Bao_Altman,Lee_Quantum_Criticality,Lee_Exact_Diagonalization,Lee_CSS_Codes,Error_Threshold_General_Stat_Mech_Mappings,Intrinsic_Error_Thresholds,Tapestry_of_dualities}. This reflects a fundamental change in the mixed state order of the system, that is independent of any specific decoder. 


For pure states, two states are said to be in the same phase if their parent Hamiltonians can be adiabatically connected along a continuous, finite path in parameter space without encountering a divergent correlation length. This is equivalent to the existence of a finite-depth (quasi)-local unitary (FDLU) circuit that maps one state to the other~\cite{Hastings_Wen_FDLU,Chen_Gu_Wen_FDLU1,Schuch_Garcia_Cirac_FDLU,Chen_Gu_Wen_FDLU2}.  
For mixed states, the absence of a similar notion of a parent Hamiltonian renders an analogous classification scheme far less straightforward. Nevertheless, in recent years, there have been various attempts to provide a comprehensive classification of such phases~\cite{Hastings_FDLC,Vidal_TEN,Castelnovo_negativity,TO_finite_temp,Negativity_Sudden_Death,coQCMI_TO_diagnostic,SPT_Order_Entangling_Surface,Classification_Mixed_State_Phaes_via_Dissipation,Sohal_Intrinsically_Mixed_State_TO, Error_Field_Double_Formalism, Markov_Length,Wang_Wu_Intrinsic_Mixed_State_TO,Sohal_Intrinsically_Mixed_State_TO,Classification_Mixed_State_TO,Sang_Mixed_State_Quantum_Phases,Anomaly_Open_Quantum_Systems,Separability_Transitions,Symmetry_Enforced_Separability_Transitions,Unconventional_Topological_Mixed_State_Transitions, Markov_Length_Local_Recovery, Diverging_Markov_Length, Lee_Bootstrap_Mixed_State}, and one of the key objects that has emerged in this discussion is the \emph{Markov length}. It is the characteristic length scale governing the decay of the conditional mutual information (CMI), just as the correlation length is the characteristic length scale governing the decay of the mutual information (MI) (see \figref{fig:Summary_Figure}). 
The operational significance of the Markov length stems from the fact that any finite depth local quantum channel that preserves the finiteness of the Markov length is (quasi)-locally reversible~\cite{Markov_Length, Markov_Length_Local_Recovery}. Thus, the Markov length is the natural analog of the correlation length for classifying mixed-state phases; it is expected to be finite within each phase and diverge only at criticality.

In pure-state criticality, both the correlation length and the Markov length diverge together. However, in mixed states, these two length scales need not track each other, opening up a new possibility: a phase with a diverging Markov length but a finite correlation length. In this work, we broaden the landscape of mixed-state phases by introducing the notion of an \emph{information critical phase}, defined as an extended region of a mixed-state phase diagram with precisely this separation of scales.

As a prototypical example, we examine two-dimensional $\mathbb{Z}_{N}$ toric codes subject to $X$ and $Z$ decoherence channels. Unlike the $\mathbb{Z}_{2}$ toric code, which exhibits a single transition from a decodable long-range-entangled phase to a non-decodable short-range-entangled phase, we find that for $N>4$, the $\mathbb{Z}_{N}$ toric code can host three distinct phases. In addition to the familiar decodable and non-decodable phases, we obtain an information critical phase in which the Markov length diverges throughout while the correlation length remains finite.

\subsection{Summary of Results}
Here we provide a brief summary of our main results. 

\begin{itemize}[leftmargin=13pt]
\item \textbf{Statistical-mechanical mapping.} In~\secref{sec:Stat_Mech}, we use analytic diagonalization techniques~\cite{Lee_Quantum_Criticality,Lee_Exact_Diagonalization,Lee_CSS_Codes,KTphase_ZdtoricCode} to diagonalize the decohered density matrices of the $\mathbb{Z}_{N}$ toric code, expressing both the coherent information $I^{(c)}$ and the CMI in terms of partition functions of a two-dimensional phase-disordered $\mathbb{Z}_{N}$ clock model. Each takes the form of a Nishimori-line disorder average of nonlinear functions of these partition functions. Consequently, the distinct thermodynamic phases of the disordered clock model correspond to different mixed-state phases of the decohered toric code. 

\item \textbf{Information critical phase.} 
In~\secref{sec:critical}, we locate an information critical phase, with divergent Markov length, in the decohered $\mathbb{Z}_N$ toric code for $N\,{>}\,4$. This phase corresponds to the quasi-long-range-ordered (QLRO) phase of the disordered clock model, which appears along the Nishimori line for $N>4$~\cite{ZN_QLRO_Existence_RG,ZN_QLRO_Existence,ZN_QLRO_Existence2,ZN_CFT_Analysis,Zn_Gauge_Glass_Nishimori_Analytics,Zn_Gauge_Glass_Nishimori_Numerics}. 
In~\appref{app:CMI_argument} we use a
Gaussian spin-wave ansatz to argue that a divergent correlation length in the
statistical-mechanics model immediately implies a divergent Markov length. 
In \secref{sec:dualSWSSB}, we further show that the R\'enyi-1 correlator of the Kramers--Wannier dual model exhibits power-law correlations throughout this regime. Since this correlator lower-bounds the CMI~\cite{Strong_to_weak_Fidelity_Correlator, Strong_Weak_SSB_Renyi1_Correlator}, $\xi_M$ diverges in the dual model as well.

\item \textbf{Fractional quantum memory.} In~\secref{sec:critical}, we also analyze the behavior of the coherent information across the decodable, non-decodable, and information critical phases. In the information critical phase, $I^{(c)}$ saturates to a fractional value in the thermodynamic limit, indicating that a fraction of the logical subspace remains protected. This is explained by a key observation: the probability of incurring a logical error is independent of system size in this phase, but remains exponentially suppressed in the $\mathbb{Z}_{N}$ charge. Thus, information stored in superpositions of logical states with large charge separation remains effectively protected, while the information stored in superpositions with small charge separation is destroyed. We further find that the two critical points bounding the phase obey a simple information-theoretic self-duality relation, given in~\eqnref{eq:self_duality}, and that the second critical point, which marks the transition to the non-decodable phase, is independent of $N$.

\item \textbf{Ensemble of quantum critical pure states.} In~\secref{sec:separability}, we use the \emph{minimally entangled typical thermal state} (METTS) ansatz~\cite{METTS,TO_finite_temp,Separability_Transitions} to expand the decohered density matrix into a convex ensemble of pure states. In the information critical phase, these states exhibit a topological tunneling amplitude that is independent of system size yet exponentially suppressed in the $\mathbb{Z}_{N}$ charge. They further display power-law correlations of string operators, allowing us to view them as Coulombic states with emergent gapless photons. The information critical phase can therefore itself be regarded as an ensemble of quantum critical pure states.

\item \textbf{Dual strong-to-weak SSB.} 
In \secref{sec:dualSWSSB}, we also consider the mixed state phase diagram of the ``ungauged" $\mathbb{Z}_{N}$ toric code obtained via Kramers-Wannier (KW) duality. Under the duality, the phase diagram can be naturally understood in the language of strong-to-weak spontaneous symmetry breaking (SWSSB)~\cite{Lee_Quantum_Criticality,Olumakinde2023,Strong_Weak_SSB_Open_Systems,Strong_to_weak_Fidelity_Correlator,Strong_Weak_SSB_Renyi1_Correlator,Strong_Weak_SSB_Hydrodynamics,Strong_Weak_SSB_SYK,Strong_Weak_SSB_Wightman,Strong_Weak_SSB_1form}.
In the KW dual model, increasing the decoherence strength causes the strong $\mathbb{Z}_{N}$ symmetry to spontaneously break to a weak $\mathbb{Z}_{N}$ symmetry. For $N>4$, an intermediate information critical phase emerges where the broken off-diagonal $\mathbb{Z}_{N}$ symmetry is effectively enhanced to a continuous $U(1)$ symmetry, giving rise to new gapless modes involving collective excitations of both the system and the environment. This yields a novel superfluid phase in these decohered discrete spin systems. 
We emphasize that the $\mathbb{Z}_N$ symmetry is inherited from $\mathbb{Z}_N$ anyon charge conservation in the original topological model. 
Although the net anyon charge remains fixed (in particular, zero) even in the presence of decoherence, sufficiently strong noise can proliferate anyon pairs, triggering a SWSSB.
 
\item \textbf{Maximum likelihood decoding.} 
Finally, in~\secref{sec:decoding}, we address the problem of decoding the noise-corrupted $\mathbb{Z}_{N}$ toric code~\cite{Dennis_2002_Error_Threshold,RG_decoder_hard_decisions,RG_Decoders_Zn_hard_decisions,RG_Decoder_soft_decisions1,RG_decoder_soft_decisions2,RG_Decoders_Zn_soft_decisions, RG_Decoders_Zn_Watson}. We implement maximum-likelihood decoding using a Monte Carlo worm algorithm combined with parallel tempering. We find that the resulting decoder exhibits error thresholds that coincide precisely with the critical points identified from the coherent information, thereby demonstrating its optimality. Within the information critical phase, the logical error rate saturates to a nonzero fractional value in the thermodynamic limit, providing an operational signature of a fractional quantum memory.
\end{itemize}

Our results uncover a new gapless analog of mixed state phases that still functions as a fractional topological memory, thereby extending the landscape of quantum memory phases beyond the conventional gapped paradigm.

\section{Preliminaries}\label{sec:Stat_Mech}
In this section, we review the key concepts and tools needed for the remainder of the paper. Readers already familiar with this material may wish to skip the corresponding subsections. In \secref{sec:toric}, we introduce the $\mathbb{Z}_N$ generalization of the toric code and define the anyon basis that underpins our subsequent analysis. In \secref{sec:quditdec}, we discuss the generalization of Pauli $X$- and $Z$-dephasing channels to $N$-dimensional qudits. In \secref{sec:Information_Diagnostics}, we review the information-theoretic quantities that serve as our primary diagnostics throughout this work.
In \secref{sec:diag}, we analytically diagonalize decohered density matrices and express all information-theoretic quantities in terms of partition functions of a disordered statistical-mechanics model. Finally, in \secref{sec:RBCM}, we provide a brief review of random bond clock models.

\subsection{$\mathbb{Z}_{N}$ toric code} \label{sec:toric}

\begin{figure}[!t]
    \centering
    \includegraphics[width=0.98\columnwidth]{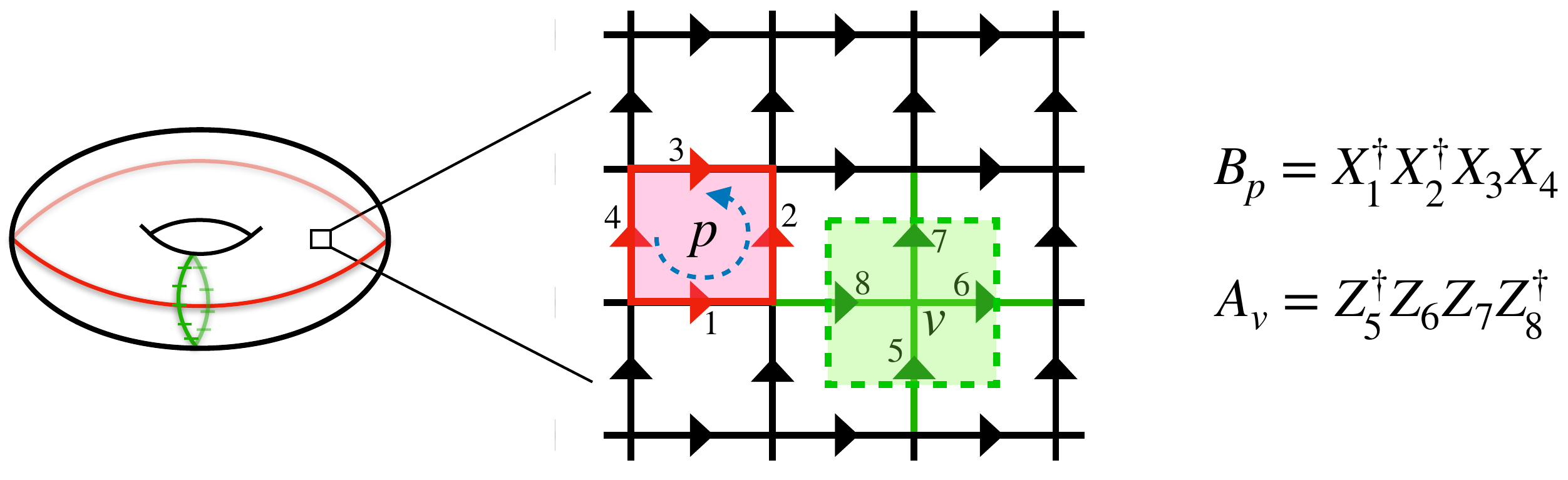}
    \caption{The plaquette and vertex stabilizers of the $\mathbb{Z}_{N}$ toric code.}
    \label{fig:Stabilizers}
\end{figure}

A $\mathbb{Z}_N$ generalization of the toric code is obtained by placing $N$-dimensional qudits on the edges of an $L \times L$ square lattice with periodic boundary conditions in both the $\hat{x}$ and $\hat{y}$ directions. Each qudit Hilbert space hosts a $\mathbb{Z}_{N}$ Pauli algebra which is generated by unitary operators $X$ and $Z$, obeying $X^{N} = Z^{N} = 1$ and $ZX = \omega XZ$, where $\omega =  e^{2\pi i/N}$ is the primitive $N$-th root of unity. We equip each link $\mu$ with an orientation as shown in \figref{fig:Stabilizers}: horizontal links point along $+\hat{x}$, and vertical links point along $+\hat{y}$.


\vspace{5pt} \noindent {\bf Stabilizers.} 
The stabilizer operators of the code are geometrically local and come in two kinds: Plaquette stabilizers $B_{p}$ and vertex stabilizers $A_{v}$.
\begin{align}
    A_{v} &= \prod_{\substack{\text{tail($\mu$)=$v$}}} Z_{\mu} \prod_{\substack{\text{ head($\mu$)=$v$}}} Z_{\mu}^{\dagger} \\
     B_{p} &=\prod_{\substack{\mu\in \partial p \\ \circlearrowleft}}X_{\mu} \prod_{\substack{\mu\in \partial p \\ \circlearrowright}}X^{\dagger}_{\mu}.
\end{align}
By convention, we choose each plaquette $p$ to be counter-clockwise oriented. It is straightforward to verify that all stabilizers mutually commute.

The logical space is defined by the condition: $A_{v}\ket{\psi} = B_{p} \ket{\psi} = \ket{\psi}$ for all vertices $v$ and all plaquettes $p$. Equivalently, it is the ground state subspace of the Hamiltonian
\begin{align}\label{eq:Hamiltonian_TC}
    H = -\sum_{v}\bigg(\frac{A_{v} + A_{v}^{\dagger}}{2}\bigg) - \sum_{p}\bigg(\frac{B_{p} + B_{p}^{\dagger}}{2}\bigg).
\end{align}
Although there are $L^2$ vertex stabilizers and $L^2$ plaquette stabilizers, they satisfy the global constraints, $\prod_{v}A_{v} = \prod_{p}B_{p} = 1$, owing to the fact that the above products always involve both $X_{\mu}$ ($Z_{\mu}$) and $X^{\dagger}_{\mu}$ ($Z^{\dagger}_{\mu}$) for each link $\mu$. Consequently, only $2L^2-2$ stabilizer constraints are independent, which means that the logical subspace consists of $2L^{2} - (2L^{2} - 2) = 2$ qudits. These logical degrees of freedom are acted upon by non-contractible Wilson and 't Hooft loop operators winding around the two independent cycles of the torus.

\vspace{5pt} \noindent {\bf Wilson and 't Hooft Loops.} 
Wilson strings reside on the direct lattice and are built from products of Pauli $X$ operators. For an oriented direct-lattice curve $\gamma$, the Wilson string that creates \emph{electric charges} $\pm k$ (violations of the vertex stabilizer condition as $A_{v} = e^{\pm 2\pi ik/N}$) at the two ends of $\gamma$ is
\begin{align}
    X_{\gamma}^{k} = \prod_{\mu \in \gamma} X^{k}_{\mu},
\end{align}
where a link $\mu$ contributes $X^{k}_{\mu}$ if it is oriented in the same sense as $\gamma$ and $(X^{k}_{\mu})^{\dagger}$ if oriented oppositely.

Conversely, 't Hooft strings reside on the dual lattice and are built from products of Pauli $Z$ operators on links intersecting the dual path. For an oriented dual lattice curve $\hat{\gamma}$, the 't Hooft string that creates \emph{magnetic fluxes} $\pm k$ (violations of the plaquette stabilizer condition as $B_{p} = e^{\pm 2\pi ik/N}$) at the 2 ends of $\hat{\gamma}$ is
\begin{align}
    Z_{\hat{\gamma}}^{k} = \prod_{\mu \perp \hat{\gamma}} Z^{k}_{\mu},
\end{align}
with the convention that a link $\mu$ contributes $Z_\mu^k$ if it crosses $\hat{\gamma}$ from left to right and $(Z_\mu^k)^\dagger$ if it crosses from right to left.

The electric and magnetic excitations of the $\mathbb{Z}_{N}$ toric code exhibit nontrivial mutual braiding statistics and therefore constitute \emph{Abelian anyons}. Closed Wilson and 't Hooft loops create no anyonic excitations and hence preserve the logical subspace. If such a loop is contractible, it can be expressed as a product of stabilizers and therefore acts trivially within the logical subspace. By contrast, non-contractible Wilson and 't Hooft loops cannot be written as products of stabilizers and thus act nontrivially on the logical degrees of freedom.

Let $\mathcal{C}_{1}$ and $\mathcal{C}_{2}$ denote non-contractible direct-lattice loops winding around the two independent cycles of the torus, with associated charge-$1$ Wilson loop operators $X_{\mathcal{C}_{1}}$ and $X_{\mathcal{C}_{2}}$. Similarly, let $\hat{\mathcal{C}}_{1}$ and $\hat{\mathcal{C}}_{2}$ denote the non-contractible dual-lattice loops, supporting the flux-$1$ 't Hooft loop operators $Z_{\hat{\mathcal{C}}_{1}}$ and $Z_{\hat{\mathcal{C}}_{2}}$. The pairs $\{X_{\mathcal{C}_{1}},Z_{\hat{\mathcal{C}}_{2}}\}$ and $\{X_{\mathcal{C}_{2}},Z_{\hat{\mathcal{C}}_{1}}\}$ obey the $\mathbb{Z}_N$ Pauli commutation relations and thus act as logical Pauli operators for the two encoded qudits.

\vspace{5pt} \noindent {\bf Logical States.} 
The logical states of the $\mathbb{Z}_{N}$ toric code are coherent superpositions of all closed, contractible string-net configurations. For instance, the logical state $\ket{\Psi_{0}}$ with eigenvalues $(Z_{\hat{\mathcal{C}}_{2}},Z_{\hat{\mathcal{C}}_{1}}) = (1,1)$ can be written as
\begin{align}\label{eq:Ground_State}
    \ket{\Psi_0} = \frac{1}{N^{(L^{2}+1)/2}}\sum_{ \{\theta_{p} = 0\cdots N-1\}} \prod_{p}B_{p}^{\theta_{p}}\ket{\{ Z_\mu = +1\}},
\end{align}
where $B_p^{\theta_p}$ creates a closed loop of charge $\theta_p$ around plaquette $p$. \eqnref{eq:Ground_State} thus realizes $\ket{\Psi_0}$ as an equal-weight superposition of all closed, contractible string-net states.
The remaining logical states are obtained by acting on $\ket{\Psi_0}$ with Wilson loop operators $X^{a_{1}}_{\mathcal{C}_{1}}$ and $X^{a_{2}}_{\mathcal{C}_{2}}$, where $a_{1},a_{2} \in \{0,\cdots, N-1\}$. These operators insert electric strings winding around the two non-contractible cycles, shifting the eigenvalues of $(Z_{\hat{\mathcal{C}}_{2}},Z_{\hat{\mathcal{C}}_{1}})$ to $(e^{2\pi i a_{1}/N},e^{2\pi i a_{2}/N})$. The logical Hilbert space therefore admits an orthonormal basis $\ket{\mathbf a} = \ket{a_{1},a_{2}}$, where $a_{1(2)}$ labels the eigenvalues of $Z_{\hat{\mathcal C}_{2(1)}}$.

\vspace{5pt} \noindent {\bf Anyon Basis.} 
A complete set of eigenstates of the Hamiltonian in \eqnref{eq:Hamiltonian_TC} is specified by the eigenvalues of the vertex stabilizers ${A_v}$, the plaquette stabilizers ${B_p}$, and the logical operators ${Z_{\hat{\mathcal C}_i}}$. Denoting these eigenvalues by $\mathbf e=\{e_v\}$, $\mathbf m=\{m_p\}$, and $\mathbf{a}$ respectively, we denote the  resulting anyon basis of the $\mathbb{Z}_{N}$ toric code as $\ket{\mathbf e;\mathbf m;\mathbf a}$.
Starting from $\ket{\Psi_0}$ in \eqnref{eq:Ground_State}, we build $\ket{\textbf{e};\textbf{m};  \textbf{a}}$ by acting with Wilson and 't Hooft strings that create the charges $\textbf{e}$ and fluxes $\textbf{m}$, together with non-contractible Wilson loops that realize the logical eigenvalues $\mathbf a$.
To describe these string configurations, we introduce $\mathbb{Z}_N$-valued flows $\textbf{k}^{e}$ and $\textbf{k}^{m}$ on the links of the lattice. The electric charges $e_v$ appear as sources of $\textbf{k}^e$, while the magnetic fluxes $m_p$ appear as vortices of $\textbf{k}^m$:
\begin{align}\label{eq:divergence_conditions}
    (\nabla\cdot \textbf{k}^e)_{v} &\equiv \sum_{\substack{\mu \\ \text{ tail($\mu$)=$v$}}}  \textbf{k}^e_{\mu} -\sum_{\substack{\mu \\ \text{ head($\mu$)=$v$}}} \textbf{k}^e_{\mu} = e_v  \mod N, \nonumber \\
    (\nabla\times \textbf{k}^m)_{p} &\equiv \sum_{\substack{\mu\in \partial p \\ \circlearrowleft}}\textbf{k}^m_{\mu}- \sum_{\substack{\mu\in \partial p \\ \circlearrowright}}\textbf{k}^m_{\mu} = m_p \mod N.
\end{align}
For non-contractible direct- and dual-lattice cycles $\mathcal{C}$ and $\hat{\mathcal{C}}$, we also define the net flow threading each cycle,
\begin{align}\label{eq:winding_def}
    \hat{\mathcal{C}}(\textbf{k}^e) = \sum_{\mu \perp  \hat{\mathcal{C}}} k^{e}_{\mu} \mod N, \\
    \mathcal{C}(\textbf{k}^m) = \sum_{\mu \in  \mathcal{C}} k^{m}_{\mu} \mod N.
\end{align}
In particular, the electric winding numbers $\hat{\mathcal{C}}_i(\textbf{k}^e)$ fix the logical eigenvalues $\textbf{a}$:
\begin{align}\label{eq:topology_condition}
    \hat{\mathcal{C}}_i(\textbf{k}^e) = a_i \iff Z_{\hat{\mathcal{C}}_{2(1)}}\big[X^{\textbf{k}^e}\ket{\Psi_0}\big] = \omega^{a_{1(2)}}X^{\textbf{k}^e}\ket{\Psi_0}.
\end{align}
The anyon-basis state can then be written compactly as
\begin{align}\label{eq:Reference_State}
    \ket{\textbf{e};\textbf{m};  \textbf{a}} = Z^{\textbf{k}^m}X^{\textbf{k}^e} \ket{\Psi_0},
\end{align}
where $X^{\textbf{k}^e}= \prod_{\mu} X^{k_{\mu}^{e}}_{\mu}$ and $Z^{\textbf{k}^m} = \prod_{\mu} Z^{k_{\mu}^{m}}_{\mu}$ denote the products of Wilson and 't Hooft string operators, respectively. Any set of flows satisfying \eqnref{eq:divergence_conditions} and $\hat{\mathcal{C}}(\textbf{k}^e) = \textbf{a}$ yield the same state, since different choices differ only by products of stabilizers, which act trivially on $\ket{\Psi_0}$.

\subsection{Qudit Decoherence Channels} \label{sec:quditdec}


For qudits, a generic $X$- or $Z$-type dephasing channel takes the form
\begin{align}\label{eq:Decoherence_Channels_ZN}
    \mathcal{E}^{x}(\rho) &= \sum_{k=0}^{N-1} p^{x}_{k}X^{k}\rho (X^{\dagger})^{k},     \\
    \mathcal{E}^{z}(\rho) &= \sum_{k=0}^{N-1} p^{z}_{k}Z^{k}\rho (Z^{\dagger})^{k}.
\end{align}
Each channel is specified by $N-1$ independent real parameters $p^{x,z}_{k} \geq 0$, subject to the normalization $\sum_{k=0}^{N-1}p^{x,z}_{k}\,{=}\,1$. We restrict to dephasing channels that preserve \emph{charge conjugation symmetry}: $p^{x,z}_k = p^{x,z}_{N-k}$. 

In the following sections, we show that the spectrum of the decohered density matrix can be expressed in terms of partition functions of two-dimensional disordered spin models whose symmetries are inherited from those of the decoherence channel. Only the long-distance behavior is relevant for our purposes, so the mixed-state phase structure is governed by the symmetry class of the associated statistical-mechanics model rather than by the microscopic values of the parameters $\{p^{x,z}_k\}$. In particular, for charge-conjugation-symmetric dephasing channels the relevant symmetry group is the dihedral group $D_N=\mathbb{Z}_N\rtimes\mathbb{Z}_2^{P}$, and the corresponding stat-mech models are non-chiral clock models. Different choices of $\{p^{x,z}_k\}$ shift only non-universal details and leave the qualitative phase diagram unchanged.

\subsection{Information Diagnostics}\label{sec:Information_Diagnostics}

In this section, we introduce the key information measures that will play a central role in the rest of the paper. Throughout this section, we shall view the $\mathbb{Z}_{N}$ toric code $Q$ as a quantum memory that encodes two qudits' worth of information in its logical space. The code is then subject to a noisy channel $\mathcal{E}$ which factorizes over physical links, $\mathcal{E} = \bigotimes_{\mu}\mathcal{E}_{\mu}$. We focus on measures that quantify the robustness of the encoded logical information and  probe the local structure of correlations.

 \vspace{5pt} \noindent {\bf Coherent Information.} 
Coherent information is the currency of quantum error correction. It upper bounds the amount of quantum information that can ultimately be recovered following the application of the noisy channel~\cite{Lloyd_Coherent_Information, Shor_Coherent_Information, Devetak_Coherent_Information, Nielsen_Chuang, Steane_bounds}. To mathematically define it, we introduce a reference system $R$ consisting of 2 qudits which are maximally entangled with the 2 logical qudits of the $\mathbb{Z}_{N}$ toric code $Q$. Together they are in the Bell state,
\begin{align}
    \ket{\Psi}_{QR} = \frac{1}{\sqrt{N^{2}}}\sum_{\textbf{a}} X^{\textbf{a}}\ket{\Psi_{0}}_{Q}\otimes\ket{\textbf{a}}_{R},
\end{align}
where $X^{\textbf{a}} \equiv X^{a_{1}}_{\mathcal{C}_{2}}X^{a_{2}}_{\mathcal{C}_{1}}$ and $\ket{\textbf{a}}_{R}=\ket{a_1,a_2}_{R}$ with $a_{i} \in \{0,\cdots,N-1\}$. The density matrix of the composite $QR$ system is $\rho_{QR} = |\Psi\rangle_{QR} \langle\Psi |_{QR}$, whereas the reduced density matrix of $Q$ is a classical mixture of all logical states,
\begin{align}\label{eq:Initial_density_matrix}
    \rho_{Q} =\frac{1}{N^{2}} \sum_{\textbf{a}}X^{\textbf{a}}|{\Psi_{0}}\rangle_{Q} \langle \Psi_{0}|_{Q} (X^{\textbf{a}})^{\dagger}.
\end{align}
The coherent information between the reference qudits $R$ and the toric code $Q$ following the application of a quantum channel $\mathcal{E}$, which acts solely on $Q$, is given by 
\begin{align}
    I^{(c)}(Q:R;\mathcal{E}) = S(\mathcal{E}(\rho_{Q})) -  S( \mathcal{E}(\rho_{QR})).
\end{align}
$I^{(c)}(Q:R;\mathcal{E})$ bounds the amount of information that can be retrieved from $\mathcal{E}(\rho_Q)$. Since the coherent information obeys the data processing inequality, $I^{(c)}(Q:R;\mathcal{E}_{1}\circ \mathcal{E}_{2}) \leq I^{(c)}(Q:R;\mathcal{E}_{2})$, information lost due to noise can never be restored by post-processing. It is for this reason that $I^{(c)}(Q:R;\mathcal{E})$ gives an algorithm-independent upper bound on recoverable information. As a consequence of the sub-additivity of entropy, it is easy to see that
\begin{align}
    -2\log N \leq I^{(c)}(Q:R;\mathcal{E}) \leq 2\log N.
\end{align}
Positive coherent information heralds distillable entanglement, whereas negative coherent information points to consumption of entanglement by the noisy channel. A negative value for the coherent information does not by itself preclude the recovery of classical information encoded in the state. 

 \vspace{5pt} \noindent {\bf Conditional Mutual Information.}
Let $A$ denote a spatial subregion and let $\rho_{A} = \Tr_{\overline{A}}{\rho}$ be its reduced density matrix. The von-Neumann entanglement entropy is given by 
\begin{align}
    S(A) = -\tr \rho_{A}\log \rho_A.
\end{align}
For a tripartite state $\rho_{ABC}$, the quantum conditional mutual information (CMI) is given by
\begin{equation} \label{eq:CMI}
    I(A:C|B)=S(AB)+S(BC)-S(B)-S(ABC).
\end{equation}
It quantifies the residual correlations (classical + quantum) between $A$ and $C$ that survive once the buffer region $B$ is known. By strong subadditivity, the CMI is non-negative and obeys the data-processing inequality. Moreover, $I(A:C\mid B) = 0$ iff $\rho_{ABC}$ is a quantum Markov state, i.e.\ it admits a direct-sum decomposition of the form
\begin{align}
    \rho_{ABC} = \bigoplus_{i}p_{i}\rho_{AB_{i}^{L}}\otimes \rho_{B_{i}^{R}C},
\end{align}
where the Hilbert space of the buffer region $B$ also decomposes as $\mathcal{H}_{B} = \bigoplus_{i}\mathcal{H}_{B_{i}^{L}}\otimes \mathcal{H}_{B_{i}^{R}}$. In this case, the tripartite state $\rho_{ABC}$ can be exactly recovered from the marginals $\rho_{AB}$ or $\rho_{BC}$ via the Petz recovery map~\cite{Quantum_Markov_Chains}. More generally, if $I(A:C|B)\leq \epsilon$ for some small, nonzero $\epsilon$, there exist recovery channels $\mathcal{R}_{B\rightarrow BC}$ (and $\mathcal{R}_{B\rightarrow AB}$) (e.g. the twirled-Petz maps) for which the reconstructed states $\mathcal{R}_{B\rightarrow BC}(\rho_{AB})$ and $\mathcal{R}_{B\rightarrow AB}(\rho_{BC})$ approach $\rho_{ABC}$ with high fidelity~\cite{Approximate_Markov_Chains}, 
\begin{align}
    F(\rho_{ABC}, \mathcal{R}_{B\rightarrow BC}(\rho_{AB})) &\geq 2^{-\epsilon/2}, \\
    F(\rho_{ABC}, \mathcal{R}_{B\rightarrow AB}(\rho_{BC})) &\geq 2^{-\epsilon/2}.
\end{align}
In this sense, the CMI quantifies approximate recoverability of $\rho_{ABC}$ from its marginals $\rho_{AB}$ or $\rho_{BC}$. 

We shall be interested in the CMI for the following geometric partition:
\begin{align}\label{eq:QCMI_geometry}
    \vcenter{\hbox{\includegraphics[height=15ex]{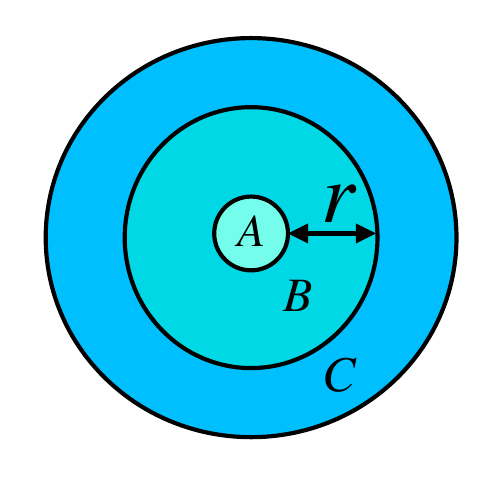}}} 
\end{align}
In non-critical models, the CMI for this geometry is expected to decay exponentially with the width of $B$,
\begin{align}
    I(A:C|B) \sim \text{poly}(|A|,|C|)  e^{-r/\xi_M}.
\end{align}
where $\xi_M$ is the \textit{Markov Length}~\cite{Markov_Length,Markov_Length_Local_Recovery, Diverging_Markov_Length}. When $\xi_M$ is finite, a small buffer region $B$ suffices to make $I(A:C|B) \approx 0$, implying conditional independence and approximate recoverability of $\rho_{ABC}$ from $\rho_{AB}$. By contrast, When $\xi_M = \infty$, approximate recovery requires a buffer whose width grows with the size of $A$, reflecting the presence of long range correlations. 
The Markov length may thus be regarded as a \textit{conditional correlation length}, measuring the spatial locality of correlations between $A$ and $C$ upon conditioning on the buffer region $B$. Crucially, if $\xi_M$ remains finite along an evolution generated by a local Lindbladian, one can construct a quasi-local quantum channel that approximately inverts the evolution, implying that all intermediate states remain in the same mixed-state phase~\cite{Markov_Length}. This is the main operational significance of the Markov length, and in this sense it plays the same role in classifying mixed-state phases that the ordinary correlation length plays in classifying pure-state phases. A divergent Markov length signals mixed-state criticality.

We emphasize that for the decohered $\mathbb{Z}_{N}$ toric code the ordinary correlation length vanishes ($\xi=0$) throughout the phase diagram. This is because the state remains a stabilizer mixture that is diagonal in the anyon-charge basis
(see~\secref{sec:diag}), so connected correlators of local operators factorize exactly and $I(A\!:\!C)$ carries no spatial tail. Nontrivial long-range structure nonetheless persists, encoded not in local correlations but in the CMI.

\subsection{Analytic Diagonalization} \label{sec:diag}

We now analytically diagonalize the decohered density matrix of the $\mathbb{Z}_{N}$ toric code in order to compute the information-theoretic quantities introduced above. Let $\rho_{Q}$ be the initial reduced density matrix of the toric code, as defined in~\eqnref{eq:Initial_density_matrix}. We subject each physical qudit to independent $X$ and $Z$ dephasing channels, specified in \eqnref{eq:Decoherence_Channels_ZN}, with channel parameters $\{ p^{x,z}_{k}\}$ taken to be identical across all qudits. The resulting post-decoherence density matrix is 
\begin{align}
    \mathcal{E}(\rho_{Q}) =  \bigg( \prod_\mu \mathcal{E}^{x}_{\mu}\circ  \prod_\mu \mathcal{E}^{z}_{\mu}\bigg)(\rho_{Q}),
\end{align}
where $\mu$ labels the links of the lattice. A key feature of the above channel is that $\mathcal{E}(\rho_{Q})$ commutes with all the stabilizer operators $A_{v}$ and $B_{p}$: 
\begin{align}
    A_{v}\mathcal{E}^{x,z}_{\mu}(\rho_{Q})  &= \mathcal{E}^{x,z}_{\mu}(\rho_{Q})A_{v}, \\
    B_{p}\mathcal{E}^{x,z}_{\mu}(\rho_{Q})  &= \mathcal{E}^{x,z}_{\mu}(\rho_{Q})B_{p}.
\end{align}
Thus $\mathcal{E}(\rho_{Q})$ can be simultaneously diagonalized with the full stabilizer algebra. Note, however, that $\mathcal{E}^{z}_{\mu}(\rho_{Q})$ and $\mathcal{E}^{x}_{\mu}(\rho_{Q})$ need not commute with the logical operators $X^{a_{1}}_{\mathcal{C}_{1}}$ and $X^{a_{2}}_{\mathcal{C}_{2}}$. In~\appref{app:diagonalizeQ}, we show that $\mathcal{E}(\rho_{Q})$ can be expanded as follows  
\begin{align}
    \mathcal{E}(\rho_{Q}) = \sum_{\substack{\textbf{e};\textbf{m} \\ \textbf{a}}} \frac{1}{N^{2}} \mathcal{Z}[\textbf{e}] \mathcal{Z}[\textbf{m}]  |\textbf{e};\textbf{m};  \textbf{a}\rangle \langle \textbf{e};\textbf{m};  \textbf{a}|,
\end{align}
where $\mathcal{Z}[\textbf{e},\textbf{a}]$ and $\mathcal{Z}[\textbf{m},\textbf{b}]$ are defined as 
\begin{align}
    \mathcal{Z}[\textbf{e},\textbf{a}] &= \sum_{\substack{(\nabla\cdot \textbf{k}^x)_{v} = e_v \\  \hat{\mathcal{C}}(\textbf{k}^{x}) = \textbf{a} }} \prod_{\mu}p^x_{k^{x}_{\mu}}, \\
    \mathcal{Z}[\textbf{m},\textbf{b}] &= \sum_{\substack{ (\nabla\times \textbf{k}^z)_{p} = m_p \\ \mathcal{C}(\textbf{k}^{z}) = \textbf{b} }} \prod_{\mu}p^z_{k^{z}_{\mu}} ,
\end{align}
and furthermore $\mathcal{Z}[\textbf{e}] = \sum_{\textbf{a}}\mathcal{Z}[\textbf{e},\textbf{a}]$ and $\mathcal{Z}[\textbf{m}] = \sum_{\textbf{b}}\mathcal{Z}[\textbf{m},\textbf{b}]$. Due to the normalization of $\{ p^{x,z}_{k}\}$, we have $\sum_{\textbf{e},\textbf{a}} \mathcal{Z}[\textbf{e},\textbf{a}] =\sum_{\textbf{m},\textbf{b}} \mathcal{Z}[\textbf{m},\textbf{b}]=1 $. 
In the next section, we show that these are normalized partition functions of two-dimensional disordered classical spin models. More precisely, $\mathcal{Z}[\textbf{e},\textbf{a}]$ ($\mathcal{Z}[\textbf{m},\textbf{a}]$) is the current loop representation of a disordered $\mathbb{Z}_{N}$ clock model on the direct (dual) lattice, evaluated within a fixed homology sector. The charges ($\mathbf{e}$) and fluxes ($\mathbf{m}$) play the role of frustrations in these disordered models.

Similarly in order to diagonalize $\mathcal{E}(\rho_{QR})$, it is convenient to introduce a new orthonormal basis of the toric code + reference Hilbert space which is defined as follows 
\begin{align}
    \ket{ \textbf{e};\textbf{m};  \textbf{a},\textbf{b}}_{QR} = \frac{1}{N}\sum_{\textbf{c}} \omega^{\textbf{b}\cdot \textbf{c}}\ket{ \textbf{e};\textbf{m};  \textbf{a}+\textbf{c}}_{Q}\otimes \ket{\textbf{c}}_{R} .\nonumber
\end{align}
We show in \appref{app:diagonalizeQ} that $\mathcal{E}(\rho_{QR})$ is fully diagonal in this basis,
\begin{align}
    \mathcal{E}(\rho_{QR})  = &\sum_{\textbf{a},\textbf{b}}\sum_{\substack{\textbf{e};\textbf{m} }} \mathcal{Z}[\textbf{e},\textbf{a}]\mathcal{Z}[\textbf{m},\textbf{b}]  \nonumber \\
    &| \textbf{e};\textbf{m};  \textbf{a},\textbf{b}\rangle_{QR} \langle  \textbf{e};\textbf{m};  \textbf{a},\textbf{b}|_{QR}.
\end{align}
Once again, the spectrum is determined entirely by the quenched partition functions $ \mathcal{Z}[\textbf{e},\textbf{a}]$ and $\mathcal{Z}[\textbf{m},\textbf{b}]$.  Note that the effect of $X$ and $Z$ decoherence decouples in the final expression in both $\mathcal{E}(\rho_{Q})$ and $\mathcal{E}(\rho_{QR})$.


\subsubsection{Coherent Information}

We now derive exact expressions for the information measures introduced in section (\ref{sec:Information_Diagnostics}) and express them in terms of $ \mathcal{Z}[\textbf{e},\textbf{a}]$ and $\mathcal{Z}[\textbf{m},\textbf{b}]$. 
Our analysis bypasses using Renyi proxies and instead focuses directly on quantities carrying direct operational meaning. This is useful since we do not rely on any analytic continuation.

The coherent information is a difference of two von-Neumann entropies,
\begin{align}
      I^{(c)}(Q:R;\mathcal{E}) = S(\mathcal{E}(\rho_{Q})) -  S(\mathcal{E}(\rho_{QR})).
\end{align}
Since we've diagonalised both $\mathcal{E}(\rho_{Q})$ and $\mathcal{E}(\rho_{QR})$, we can write down a simple, exact closed form expression for $I^{(c)}(Q:R;\mathcal{E})$. First note that the von-Neumann entropy of $\mathcal{E}(\rho_{Q})$ is
\begin{align}\label{eq:vN_Entropy}
    S(\mathcal{E}(\rho_{Q})) = 2\log N &- \sum_{\textbf{e}} \mathcal{Z}[\textbf{e}] \log \mathcal{Z}[\textbf{e}]  \\
    &- \sum_{\textbf{m}} \mathcal{Z}[\textbf{m}] \log \mathcal{Z}[\textbf{m}]\nonumber.
\end{align}
On the other hand the Von-Neumann entropy of $\mathcal{E}(\rho_{QR})$ is given by
\begin{align}
    S(\mathcal{E}(\rho_{QR}) ) = &-\sum_{\textbf{e}, \textbf{a}} \mathcal{Z}[\textbf{e}, \textbf{a}]\log \mathcal{Z}[\textbf{e}, \textbf{a}]\nonumber \\ &-\sum_{\textbf{m}, \textbf{b}}\mathcal{Z}[\textbf{m}, \textbf{b}]\log \mathcal{Z}[\textbf{m}, \textbf{b}].
\end{align}
Thus, the coherent information takes the simple form
\begin{align}
     I^{(c)}(Q:R;\mathcal{E}) = &2\log N +\sum_{\textbf{e}, \textbf{a}}\mathcal{Z}[\textbf{e}, \textbf{a}]\log\frac{ \mathcal{Z}[\textbf{e}, \textbf{a}]}{\sum_{\textbf{a}'} \mathcal{Z}[\textbf{e}, \textbf{a}']} \nonumber \\
     &+\sum_{\textbf{m}, \textbf{b}}\mathcal{Z}[\textbf{m}, \textbf{b}]\log\frac{ \mathcal{Z}[\textbf{m}, \textbf{b}]}{\sum_{\textbf{b}'} \mathcal{Z}[\textbf{m}, \textbf{b}']}. 
\end{align}
We can view these expressions as disorder averages over frustration variables $\textbf{e}$ and $\textbf{m}$. Since the disorder distribution is precisely given by the partition function, the model falls on the Nishimori line~\cite{Spin_Glasses_Nishimori,Yozeki_1993}. The Nishimori line is very special because the disordered model acquires an effective local $\mathbb{Z}_{N}$ symmetry. This is unsurprising in our context, since the partition functions are describing spectra of density matrices in a gauge theory. 

Furthermore, the ratios $P[\textbf{a}|\textbf{e}] = \frac{ \mathcal{Z}[\textbf{e}, \textbf{a}]}{\sum_{\textbf{a}'} \mathcal{Z}[\textbf{e}, \textbf{a}']}$ and $P[\textbf{b}|\textbf{m}] = \frac{ \mathcal{Z}[\textbf{m}, \textbf{b}]}{\sum_{\textbf{b}'} \mathcal{Z}[\textbf{m}, \textbf{b}']}$ correspond to winding number probabilities conditioned on the disorder realization. Therefore, $I^{(c)}(Q:R;\mathcal{E})$ can equivalently be expressed as
\begin{align}
    I^{(c)}(Q:R;\mathcal{E}) = 2\log N &-  [ H(\textbf{a}|\textbf{e})] - [ H(\textbf{b}|\textbf{m}) ],
\end{align}
where $H(\textbf{a}|\textbf{e})$ and $H(\textbf{b}|\textbf{m})$ denote Shannon entropies of the conditional distributions $P[\textbf{a}|\textbf{e}]$ and $P[\textbf{b}|\textbf{m}]$ respectively, and $[\cdots]$ denotes a disorder average on the Nishimori line. This final expression for the coherent information is particularly well suited for numerical evaluation.

\subsubsection{Conditional Mutual Information}

To compute the conditional mutual information, we need to work with reduced density matrices. Consider the toric code logical state $\ket{\Psi_0}$ subject to only $X$ type dephasing. In~\appref{app:CMI_appendix}, we show that the conditional mutual information for the geometry illustrated in~\eqref{eq:QCMI_geometry} reduces to the following simple form,
\begin{align}\label{eq:QCMI}
    I(A:C|B) = &- \sum_{\textbf{e}_{AB}}\mathcal{Z}[\textbf{e}_{AB}]\log \mathcal{Z}[\textbf{e}_{AB}] \nonumber \\
    &- \sum_{\textbf{e}_{BC}, E_{A}}\mathcal{Z}[\textbf{e}_{BC}, E_{A}]\log \mathcal{Z}[\textbf{e}_{BC}, E_{A}] \nonumber \\
    &+ \sum_{\textbf{e}_{B},E_{A}}\mathcal{Z}[\textbf{e}_B, E_{A}]\log \mathcal{Z}[\textbf{e}_B, E_{A}] \nonumber \\
    &+ \sum_{\textbf{e}_{ABC}}\mathcal{Z}[\textbf{e}_{ABC}]\log \mathcal{Z}[\textbf{e}_{ABC}].
\end{align}
Here $\textbf{e}_{Q} = \{ e_{v}\,|\,v\in \text{Int}(Q)\}$ denotes the electric charges in the interior of region $Q$ (excluding the boundary $\partial Q$), and $E_{Q} = \sum_{v\in \text{Int}(Q)} e_{v}$ is the total enclosed charge. The reduced partition function on $Q$ is obtained by summing over all charge configurations outside $Q$, $\mathcal{Z}[\textbf{e}_{Q}] = \sum_{\textbf{e}_{v\notin \text{Int}(Q)}}\mathcal{Z}[\textbf{e}]$. Since $B$ and $BC$ are annular, their partition functions depend on the total charge inside the hole, which is precisely $E_{A}$.

In~\appref{app:relative_entropy}, we also consider the relative entropy of two decohered logical states and express it in terms of conditional winding number distributions.

\subsection{Random Bond Clock Models} \label{sec:RBCM}

In the previous section, we expressed all relevant quantities in terms of $\mathcal{Z}[\mathbf{e},\mathbf{a}]$ and $\mathcal{Z}[\mathbf{m},\mathbf{a}]$. We now demonstrate that these quantities can be interpreted as partition functions of two-dimensional classical spin models.

We first parametrize the channel weights as $p^{x,z}_{k} = e^{\alpha^{x,z}_{k}}/\mathcal{N}_{x,z}$, with $\mathcal{N}_{x,z} = \sum_{k =0}^{N-1}e^{\alpha^{x,z}_{k}}$, and introduce the discrete Fourier transform of $\{\alpha^{x,z}_{k}\}$,
\begin{align}
    \beta^{x,z}_{m} = \frac{1}{N}\sum_{k = 0}^{N-1}\alpha^{x,z}_{k} \omega^{-km} \quad, \quad
    \alpha^{x,z}_{k} = \sum_{m = 0}^{N-1}\beta^{x,z}_{m} \omega^{km}.
\end{align}
Now consider
\begin{align}\label{eq:current_loop_partition_function}
    \mathcal{Z}[\textbf{e},\textbf{a}] &= \sum_{\substack{(\nabla\cdot \textbf{k}^x)_{v} = e_v \\  \hat{\mathcal{C}}(\textbf{k}^{x}) = \textbf{a} }} \prod_{\mu}p^x_{k^{x}_{\mu}}. 
\end{align}
We need to parameterize all flow configurations $\textbf{k}^{x}$ that solve the constraints $(\nabla\cdot \textbf{k}^x)_{v} = e_v $ and $\hat{\mathcal{C}}(\textbf{k}^{x}) = \textbf{a}$. $\textbf{k}^{x}$ consists of $2L^{2}$ qudit degrees of freedom on the links, subject to $L^{2} + 1$ constraints: $L^{2}-1$ independent divergence constraints together with two homology constraints along the non-contractible cycles of the torus. That leaves $2L^{2} - (L^{2}+1) = L^{2}-1$ independent qudit degrees of freedom. Let $k^{e}_{\mu, \textbf{a}}$ be a reference configuration which solves $(\nabla\cdot \textbf{k}^e_{\textbf{a}})_{v} = e_v $ and $\hat{\mathcal{C}}(\textbf{k}^{e}_{\textbf{a}}) = \textbf{a}$. All other allowed flow configurations are related to this reference configuration via additional closed, contractible Wilson loops, $\textbf{k}^x = \textbf{k}^e_{\textbf{a}} + \textbf{l}^x$ with $(\nabla\cdot \textbf{l}^{x})_{v} = 0 $ and $\hat{\mathcal{C}}( \textbf{l}^{x}) = 0$. A convenient way to parameterize the set of all $\textbf{l}^{x}$ configurations is to introduce $\mathbb{Z}_{N}$ spins $\theta_{p}$ at the center of each plaquette $p$ and write
\begin{align}
    l^{x}_{\mu \perp\langle p_{1}\rightarrow p_{2}\rangle} &= \theta_{p_{1}} - \theta_{p_{2}} \mod N,
\end{align}
where $\langle p_{1}\rightarrow p_{2}\rangle$ denotes the unique dual lattice link from $p_{1}$ to $p_{2}$ that intersects the direct lattice link $\mu$. We adopt the orientation convention that if a direct lattice link goes from bottom to top, the unique dual link which intersects it crosses it from left to right. Now for any configuration of spins $\{\theta_{p} \}$, we have $(\nabla\cdot \textbf{l}^x)_{v} = 0$ and $\hat{\mathcal{C}}(\textbf{l}^{x}) = 0$. $\{ l^{x}_{\mu}\}$ are domain walls between neighboring spin configurations. Note that since $l^{x}_{\mu}$ is unchanged by the global shift $\theta_{p}\rightarrow (\theta_{p} + a) \mod N$ with $a = 0,\cdots,N-1$, we recover precisely ${L^{2}-1}$ independent qudit degrees of freedom. In terms of the spins $\{\theta_p\}$, $\mathcal{Z}[\mathbf{e}, \mathbf{a}]$ is the partition function of a spin model on the dual lattice with nearest-neighbor interactions and quenched bond disorder set by $\hat{k}^{e}_{\mu;\mathbf{a}}$:
\begin{align}
    \mathcal{Z}[\textbf{e}, \textbf{a}] &= \frac{1}{\mathcal{N}^{2L^{2}}_{x}} \sum_{\{ \theta_{p}\}} \exp{-H[\{\theta_{p}\}; \textbf{k}^{e}_{\textbf{a}}]} , \\
    H[\{\theta_{p}\}; \textbf{k}^{e}_{\textbf{a}}] &= -\sum_{\langle p\rightarrow p' \rangle}{\sum_{m = 0}^{N-1}\beta^x_{m}\omega^{m(\hat{k}^{e}_{\langle p \rightarrow p' \rangle,\textbf{a}} + \theta_{ p} - \theta_{ p'} )}}.
\end{align} 
The parameters $\beta^{x}_{m}$ correspond to coupling constants that source distinct harmonic interactions in the spin model. For generic choices of ${\beta^{x}_m}$, this spin model has a sign problem. However, for charge conjugation symmetric channels, this obstruction is lifted since only cosine harmonics survive and the resulting Boltzmann weights become positive definite.
\begin{align}\label{eq:Hamiltonian_clock}
    H[\{\theta_{p}\}; \textbf{k}^{e}_{\textbf{a}}] &= -\sum_{\langle p, p'\rangle}\sum_{m = 0}^{[N/2]}  \beta^{x}_{m}\nonumber \\
    &\times\cos\bigg(\frac{2\pi m(k^{e}_{\langle p , p'\rangle,\textbf{a}} + \theta_{p} - \theta_{p'} )}{N}\bigg). 
\end{align}
This is now the Hamiltonian for a random phase clock model (a ``discrete gauge glass") with coupling constants $\beta^{x}_m$ which source different cosine harmonics. The vertex charges $e_v$ become local bond frustrations on the dual lattice $(\curl \hat{\textbf{k}}^{e}_{\textbf{a}})_{p} = e_{p} \mod N$. If $e_p \neq 0$, no configuration of the clock spins can simultaneously minimize all interactions around the corresponding plaquette, thereby signaling local frustration. Since the partition function is invariant under the ``gauge transformation" $\hat{k}^{e}_{\langle p \rightarrow p' \rangle,\textbf{a}} \rightarrow \hat{k}^{e}_{\langle p \rightarrow p' \rangle,\textbf{a}} + \alpha_{p} - \alpha_{p'}$, it can only depend on gauge invariant components of the quenched disorder, which are precisely these frustrations. Finally, the logical eigenvalues $\textbf{a} = (a_{1},a_{2})$ impose twisted boundary conditions along the non-contractible cycles of the torus. These twists introduce global, topological frustrations in the clock model, leading to different statistical weights for each winding number sector. 

An analogous construction applies to
\begin{align}
    \mathcal{Z}[\textbf{m},\textbf{a}] &= \sum_{\substack{ (\nabla\times \textbf{k}^z)_{p} = m_p \\ {\mathcal{C}}(\textbf{k}^{z}) = \textbf{a} }} \prod_{\mu}p^z_{k^{z}_{\mu}}.
\end{align}
Once again, $\mathbf{k}^z$ involves $2L^{2}$ link degrees of freedom, subject to $L^{2}+1$ constraints: $L^{2}-1$ independent curl constraints together with two homology constraints along the non-contractible cycles of the torus. This leaves $L^{2}-1$ independent qudit degrees of freedom. Let $k^{m}_{\mu;\mathbf{a}}$ denote a fixed reference configuration satisfying $(\curl \textbf{k}^m_{\textbf{a}})_{p} = m_p$ and ${\mathcal{C}}(\textbf{k}^{m}_{\textbf{a}})=\textbf{a}$. All other allowed configurations are now related to this reference configuration through the action of closed, contractible ’t Hooft loops, namely $\textbf{k}^z = \textbf{k}^m_{\textbf{a}} + \textbf{l}^z$ with $(\curl  \textbf{l}^{z})_{p} = 0 $ and ${\mathcal{C}}( \textbf{l}^{z}) = 0$. Again a convenient way to parametrize the full set of $\mathbf{k}^z$ configurations is to introduce $\mathbb{Z}_N$-valued spin variables $\theta_v$ residing on each vertex $v$, and write
\begin{align}
    l^{z}_{\mu = \langle v_{1}\rightarrow v_{2}\rangle} =  \theta_{v_{1}} - \theta_{v_{2}} \mod N.
\end{align}
For any configuration of spins $\{\theta_{v} \}$, this gives $(\curl \textbf{l}^z)_{p} = 0$ and ${\mathcal{C}}(\textbf{l}^{z}) =0$. Furthermore, since $l^{z}_{\mu}$ is unchanged by the global shift $\theta_{v}\rightarrow (\theta_{v} + b) \mod N$ with $b = 0,\cdots,N-1$, this parametrization yields precisely ${L^{2}-1}$ independent qudit degrees of freedom. In terms of spin variables $\{ {\theta_v}\}$, $\mathcal{Z}[\mathbf{m}, \mathbf{a}]$ becomes the partition function of a clock model on the direct lattice with nearest-neighbor interactions and quenched bond disorder set by the reference configuration $k^{m}_{\mu;\mathbf{a}}$,
\begin{align}
    \mathcal{Z}[\textbf{m}, \textbf{a}] &= \frac{1}{\mathcal{N}^{2L^{2}}_{z}}\sum_{\{ \theta_{v}\}} \exp{-H[\{\theta_{v}\}; \textbf{k}^{m}_{\textbf{a}}]} ,\nonumber \\
    H[\{\theta_{v}\}; \textbf{k}^{m}_{\textbf{a}}]&= -\sum_{\langle v\rightarrow v' \rangle} {\sum_{n = 0}^{N-1}\beta^z_{n}\omega^{n(k^{m}_{\langle v \rightarrow v' \rangle,\textbf{a}} + \theta_{ v} - \theta_{ v'} )}}.
\end{align} 
As before, the parameters $\beta^{z}_{n}$ act as coupling constants that source distinct harmonic interactions in the spin model. Assuming charge conjugation symmetry of the channel again reduces this to the disordered non-chiral clock model, and the magnetic fluxes $m_p$ play the role of local frustrations.

We have thus shown above that the spectra of the decohered density matrices can be expressed in terms of partition functions of 2D disordered clock models. Since the evaluation of information theoretic measures reduces to computing disorder-averaged expectation values of nonlinear functions of these partition functions along the Nishimori line, the problem of characterizing mixed-state phases of the decohered $\mathbb{Z}_{N}$ toric code essentially reduces to understanding the thermodynamic phases of these disordered clock models on the Nishimori line.

Two-dimensional $\mathbb{Z}_{N}$ clock models have been studied extensively in the literature. Notable special cases include the Ising model ($N = 2$), three-state Potts model $(N = 3)$, the Ashkin-Teller model $(N = 4)$ and the XY model ($N\rightarrow \infty$). These models possess a dihedral symmetry $D_{N} = \mathbb{Z}_{N}\rtimes\mathbb{Z}^{P}_{2}$ and, in the absence of disorder, they are known to display a rich phase diagram that changes qualitatively with increasing $N$. In particular, for $2\leq N\leq 4$, the $\mathbb{Z}_{N}$ clock model exhibits only two phases-a low-temperature ordered phase and a high-temperature disordered phase-separated by a single critical point. On the other hand, for $N> 4$, the $\mathbb{Z}_{N}$ clock model is found to exhibit three different phases: a low temperature ordered phase, a high temperature disordered phase, and an intervening quasi-long-range-ordered (QLRO) critical phase~\cite{ZN_QLRO_Existence,ZN_QLRO_Existence_RG,ZN_QLRO_Existence2,ZN_CFT_Analysis,Z6_numerics, ZN_clock_numerics, ZN_clock_numerics2, Domany_1980, Z6_Phase_Diagram, No_Disorder_Clock_model_Phases}. Within this intermediate phase, both the domain wall and vortex excitations are RG irrelevant, and the IR effective theory becomes a Gaussian spin-wave theory with an emergent $U(1)$ symmetry. Crucially, because the low-energy excitations in this phase are gapless, the correlation length diverges throughout the phase. The existence of such a QLRO phase follows from general renormalization group arguments, and it does not rely on any fine-tuned choice of coupling constants. Microscopic couplings merely affect nonuniversal properties, such as the exact locations of the critical points, but they do not alter the qualitative structure of the phase diagram.
\begin{figure*}
    \includegraphics[width=0.98\textwidth]{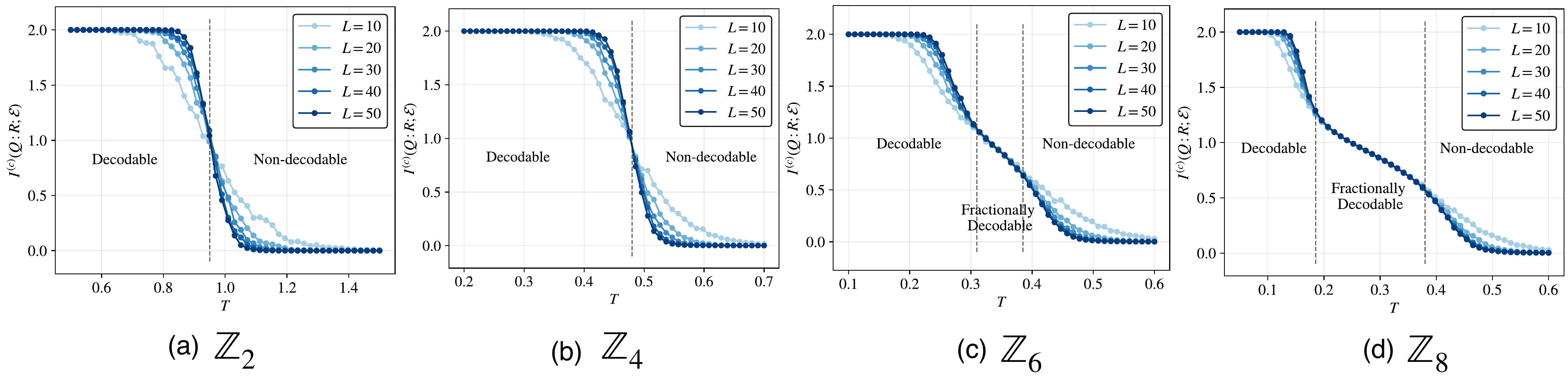}
    \caption{\label{fig:Coherent_Information}  {\bf Coherent Information.} We numerically compute the coherent information of the $\mathbb{Z}_{N}$ toric code for $N = 2, 4, 6,$ and $8$ under $X$-type dephasing. $L$ denotes the linear size of the toric code. The decoherence parameters $p^{x}_{k}$ are fixed as in~\eqnref{eq:Fixed_Parameters}, with the decoherence temperature $T \equiv T^{x}$ acting as the sole control parameter. To enable a direct comparison across different values of $N$, the vertical axis is plotted in logarithmic units of base $N$. For $N \leq 4$, we observe a single phase transition in the disordered clock model, separating an ordered and a disordered phase. In the ordered phase, the coherent information saturates to $2 \log N$, indicating perfect decodability of the encoded logical information. In contrast, in the disordered phase the coherent information vanishes in the thermodynamic limit, signaling the complete loss of encoded quantum information. For $N > 4$, however, the corresponding clock model exhibits three distinct phases: an ordered phase, a disordered phase, and an intervening QLRO phase. Remarkably, within the QLRO phase the coherent information saturates to a nonzero fractional value as $L \to \infty$, indicating that a finite fraction of logical information is still protected. The extent of this fractionally decodable phase grows with increasing $N$, and we find that the second critical temperature $T^{(2)}_c \approx 0.38$ is approximately independent of the value of $N$. Interestingly, the locations of the critical points are found to agree to excellent precision with the self-dual identity given in~\eqnref{eq:self_duality}.} 
\end{figure*}

In the presence of the quenched phase disorder, the fate of the intermediate QLRO phase becomes uncertain~\cite{QLRO_Disorder}. Indeed, it is known that sufficiently strong disorder completely destroys all signatures of the QLRO phase. However, when restricted to the Nishimori line, it is generally expected that the qualitative structure of the phase diagram mirrors that of the clean $\mathbb{Z}_{N}$ clock model~\cite{Zn_Gauge_Glass_Nishimori_Analytics,Zn_Gauge_Glass_Nishimori_Numerics}. In the next section, we verify numerically that this is indeed the case. Precisely which phase the decohered toric code occupies ultimately depends on the decoherence channel and the choice of parameters $p^{x,z}_{k}$. In the next section, we make a concrete choice for these parameters and numerically confirm the emergence of all three phases along the Nishimori line for $N>4$. We then examine the implications of these phases for the information-theoretic measures introduced above. Note that a similar analysis was carried out for measurement-induced phases in symmetric quantum circuits~\cite{vijay_holographicallyemergentgaugetheory}.

\section{Information Critical Phases} \label{sec:critical}

In this section, we define the \emph{information critical phase}, and demonstrate that such a phase arises in decohered $\mathbb{Z}_N$ toric codes for $N > 4$.

\subsection{Intrinsic mixed-state behavior}

It is instructive to compare the CMI with the ordinary mutual information (MI) between subregions $A$ and $C$,
\begin{align}
    I(A:C) := S({A}) + S({C}) - S({AC}).
\end{align}
The MI upper-bounds all connected correlators of operators supported on $A$ and $C$,
\begin{align}
    \frac{1}{2} \qty( \frac{\langle O_A O_C \rangle_c}{\Vert O_A \Vert \cdot \Vert O_C \Vert} )^2 \leq I(A:C),
\end{align}
where $\langle \cdot \rangle_c$ denotes the connected correlation function, and $\Vert O \Vert := \max_\psi \sqrt{\langle \psi | O^\dagger O | \psi \rangle }$ is an operator norm. In a short-range correlated phase, the MI typically decays with the separation $r:= \textrm{dist}(A,C)$ as
\begin{align}
    I(A:C)\sim \text{poly}(|A|,|C|)  e^{-r/\xi},
\end{align}
where $\xi$ is just the familiar correlation length. This behavior causes all connected correlators to also decay exponentially on the same length scale $\xi$. Divergence of $\xi$ signals conventional criticality, where power-law correlations may emerge. 

Interestingly, if $\xi_M = \infty$ but $\xi < \infty$, we obtain a different kind of criticality that does not require long-range two-point correlations. This separation between the behaviors of $I(A:C)$ and $I(A:C|B)$ is an \emph{intrinsically mixed-state phenomenon}. If $\rho_{ABC}$ is pure, then,
\begin{align}
    I(A:C) = I(A:C|B) \quad \textrm{(for pure states)},
\end{align}
so the correlation and Markov lengths necessarily diverge together at pure-state critical points. A separation of scales between $\xi_M$ and $\xi$ is a genuine mixed-state phenomenon. Known examples of such \emph{information criticality} are decoherence-induced critical points of the $\mathbb{Z}_{2}$ toric code where $\xi = 0$ but $\xi_M =\infty$. 

In the pure-state setting one encounters not only isolated quantum critical points but also quantum critical phases---contiguous regions of phase space with a diverging correlation length. By analogy, we define an \emph{information critical phase} as an extended region of phase space with a diverging Markov length ($\xi_M = \infty$) but a finite correlation length ($\xi < \infty$). By virtue of being an extended phase rather than an isolated point, it is stable against certain (symmetry-respecting) perturbations up to a finite threshold. While we do not assume any additional structure, such as emergent conformal invariance or a continuous manifold of renormalization-group fixed points, in this definition, it is natural to ask whether such structure nevertheless emerges. 

We now study the mixed-state phase diagram of the $\mathbb{Z}_N$ toric code under decoherence and present evidence for an information critical phase when $N>4$.

\subsection{Numerical details}\label{sec:Numerical_details}

For convenience, we take the $X$ and $Z$ decoherence parameters to be
\begin{align}\label{eq:Fixed_Parameters}
    p^{x,z}_{k} = \frac{1}{\mathcal{N}^{x,z}} \exp(\frac{1}{T^{x,z}}\cos{\frac{2\pi k}{N}}),
\end{align}
with $\mathcal{N}^{x,z}$ a normalization constant. 
The sole control parameter is then the effective \emph{decoherence temperature} $T^{x,z}$, which sets the noise strength. For small $T^{x,z}$, the distribution $p^{x,z}_{k}$ is sharply peaked at $k=0$ (weak decoherence), whereas for large $T^{x,z}$ it becomes approximately uniform, $p^{x,z}_{k} \simeq 1/N$ (strong decoherence). In the notation of~\eqnref{eq:Hamiltonian_clock}, this choice retains only the first non-trivial cosine harmonic, with coupling $\beta^{x,z}_{1}=1/T^{x,z}$, and all higher harmonics $\beta^{x,z}_{m\neq 1}$ set to zero.

We now outline the numerical procedure. Throughout, we restrict attention to $X$-type dephasing---which sources only electric-charge excitations---since $Z$-type dephasing is entirely analogous. The coherent information requires evaluating the conditional winding-number distribution $P[\textbf{a}|\textbf{e}] = \frac{ \mathcal{Z}[\textbf{e}, \textbf{a}]}{\sum_{\textbf{a}'} \mathcal{Z}[\textbf{e}, \textbf{a}']}$. We evaluate this quantity efficiently using the worm algorithm~\cite{Worm_Algorithm1,Worm_Algorithm2,Worm_Algorithm_3D_U1}, which samples flow configurations in the current-loop expansion of the disordered clock model, as given in~\eqnref{eq:current_loop_partition_function}. We parametrize the allowed flow configurations as $\mathbf{k} = \mathbf{k}' + \mathbf{l}$, where the reference $\mathbf{k}'$ satisfies the divergence constraints $(\nabla\cdot \mathbf{k}')_{v} =e_{v}$, leaving the fluctuation $\mathbf{l}$ divergence-free, $(\nabla\cdot \mathbf{l})_{v} =0$. Such divergence-free configurations are sampled by sequentially adding closed loops (``worms'') of charge: the algorithm temporarily relaxes the divergence constraint by inserting a head--tail defect pair that locally violates charge conservation, then propagates the head via local updates weighted by the modified Boltzmann factor $\prod_{\mu} p_{k'_{\mu}+l_{\mu}}$. When the head returns to the tail, the defects annihilate and the configuration again satisfies the original constraints. Ordinary expectation values are obtained by averaging over closed-worm configurations, while intermediate open-worm configurations give defect correlation functions. The worm algorithm is ergodic across all topological sectors~\cite{Worm_Algorithm1,Worm_Algorithm2, Worm_Algorithm_3D_U1}, making it well suited to computing $P[\mathbf{a}|\mathbf{e}]$ at fixed charge backgrounds. Finally, since we ultimately want disorder-averaged observables along the Nishimori line, the reference configurations $\mathbf{k}'$ may be sampled directly from the product distribution $p(\mathbf{k}')=\prod_{\mu}p_{k'_{\mu}}$.

We now discuss the resulting phase diagram of the $X$-decohered toric code, contrasting $N\leq 4$ with $N>4$.

\subsection{$N\leq4$}
For $N\leq 4$, the random bond clock model exhibits only 2 distinct phases: a low-temperature ordered phase and a high-temperature disordered phase. The Markov length remains finite in each phase, and so the CMI decays exponentially in the thickness of $B$,
\begin{align}
    I(A:C|B) \sim e^{-r/\xi_{M}(T^x)}.
\end{align}
Furthermore, in the ordered phase the probability of creating long worms that wind nontrivially around the torus is exponentially suppressed in the system size. Thus, the conditional winding-number distribution $P[\mathbf{a}|\mathbf{e}]$ is sharply peaked at $\mathbf{a}=0$, up to exponentially small corrections. This gives $[ H(\textbf{a}|\textbf{e})] \approx 0$, so the coherent information remains approximately maximal,
\begin{align}
    I^{(c)}(Q:R;\mathcal{E}) \approx 2\log N + \mathcal{O}(e^{-\alpha L}).
\end{align}
The system thus realizes a decodable phase in which logical information is asymptotically perfectly protected in the thermodynamic limit $(L\rightarrow \infty)$. On the other hand, in the disordered phase, the free energy cost of system spanning loops vanishes, leading to uniform sampling over all topological sectors. This implies that  $P[\mathbf{a}|\mathbf{e}] \approx \frac{1}{N^{2}}$, yielding,
\begin{align}
    I^{(c)}(Q:R;\mathcal{E}) \approx 0 + \mathcal{O}(e^{-\alpha' L}).
\end{align}
All logical quantum information is destroyed in this regime, and the system can function at most as a classical memory. These expectations are confirmed numerically in~\figref{fig:Coherent_Information} for the $\mathbb{Z}_{2}$ and $\mathbb{Z}_{4}$ toric codes. At the critical point, the coherent information becomes independent of system size $L$ signaling emergent scale invariance. This is a direct consequence of the fact that the disordered clock model is at criticality and exhibits power law correlations. At this point, the Markov length diverges ($\xi_{M} = \infty$) and the CMI decays only algebraically. This is a genuine mixed-state phase transition of the decohered toric code.

\subsection{$N> 4$}
For $N>4$, the ordered and disordered phases are qualitatively similar to the $N\leq 4$ case. These phases continue to correspond to the decodable and non-decodable phases respectively, both with a finite Markov length. However, in the intermediate QLRO phase, the clock model is scale-invariant with a divergent correlation length over an extended region of the phase diagram. We identify this as an information critical phase. 

In the clean clock model, the QLRO phase is governed in the infrared by a Gaussian spin-wave theory, and the two critical points are of Berezinskii–Kosterlitz–Thouless (BKT) type. In the disordered case, along the Nishimori line, we again expect the effective low-energy theory to be Gaussian, but with a stiffness renormalized by bulk disorder. Distinct topological sectors in this phase are related by harmonic zero modes winding around the torus. As argued in ~\appref{app:clock_model}, the energies of these winding modes are independent of system size, leading to conditional winding number probabilities
\begin{align}
    P(\textbf{a}|\textbf{e}) \sim e^{-\frac{\kappa(T^x)}{2} |\textbf{a}|^{2}},
\end{align}
where $\kappa(T^{x})$ is the renormalized spin stiffness, which depends on the decoherence strength, and ${\textbf{a}}$ is now shifted to the symmetric integer range $\{-\lfloor N/2\rfloor,\ldots,\lfloor N/2\rfloor\}$. Crucially, these probabilities are independent of system size yet remain exponentially suppressed in the topological charge $\textbf{a}$. These two properties translate directly into the phase's information-theoretic signatures. The first is a divergent Markov length. In~\appref{app:CMI_argument} we use a Gaussian spin-wave ansatz to argue that a divergent correlation length in the statistical-mechanics model immediately implies a divergent Markov length. 

The second signature is the behavior of the coherent information: the emergent scale invariance makes it independent of system size $L$ across the phase, so it saturates to a finite, fractional value in the thermodynamic limit. This fractional plateau signals the persistence of a partially protected logical subspace. This expectation is borne out numerically in~\figref{fig:Coherent_Information} for the $\mathbb{Z}_{6}$ and $\mathbb{Z}_{8}$ toric codes. We observe from our numerics that the second critical point $T^{(2)}_{c}\approx 0.38$ is approximately independent of $N$ just as for the clean clock model.

\subsection{Critical points and self-duality}

The robustness of logical information encoded in the $\mathbb{Z}_{N}$ toric code can be contrasted with a minimal setting involving two raw qudits subject to independent $X$ and $Z$ dephasing channels, characterized by effective temperatures $T^{x}$ and $T^{z}$. The coherent information of this two-qudit system is given by
\begin{align}
    I^{(c)}_{\text{raw}} = 2\Big(\log N - \sum_{i=x,z} H(T^{i})\Big),
\end{align}
where $H(T^{i}) = -\sum_{k = 0}^{N-1}p^{i}_{k}(T^{i})\log p^{i}_{k}(T^{i}) $ is the Shannon entropy of the distributions $p^{i}_{k}(T^{i})$ defined in~\eqnref{eq:Fixed_Parameters}. 

For the toric code, the coherent information stays at its maximal value $2\log N$ up to a finite threshold $T^{i} \leq T^{(1)}_{c}$, reflecting the intrinsic robustness afforded by topological encoding. In contrast, for the raw qudits, the coherent information degrades continuously as the noise strength $T^{i}$ is increased. For $N=2$, it is well known that the numerically observed error threshold $T_{c}$ (or equivalently $p_c$) of the toric code is, to an excellent approximation, determined by setting $T^{x} = T^{z}$ and finding the point where $I^{(c)}_{\text{raw}} = 0$. This coincides with the asymptotic Gilbert-Varshamov bound for CSS-type codes~\cite{CSS_1996,Steane_bounds}. 

We find that the same condition, $\log N = 2H(T_c)$, continues to predict the critical point $T_c$ of $I^{(c)}(Q:R;\mathcal{E})$ to excellent precision for all $N \leq 4$. More remarkably, for $N>4$---where two transitions appear at $T^{(1)}_{c}$ and $T^{(2)}_{c}$---both critical points obey the self-duality relation given by $I^{(c)}_{\text{raw}} = 0$:
\begin{align}\label{eq:self_duality}
    \log N = H(T^{(1)}_{c}) + H(T^{(2)}_{c}).
\end{align}
While we are unable to provide an information-theoretic explanation for this fact at this stage, we note that the same relation has previously appeared as a conjectured self-duality condition for disordered spin models along the Nishimori line~\cite{Duality_RBIM,Duality_RBIM_toric_Code,Duality_symmetry_spin_glasses,Duality_finite_dimensional_spin_glasses,Duality_Zn_glasses,No_spin_glass_transition_self_dual_lattices}. These observations reveal a deep link between self-duality in disordered spin systems along the Nishimori line and the error thresholds of topological quantum error-correcting codes.


\section{Separability Transitions} \label{sec:separability}
In the $\mathbb{Z}_{2}$ toric code, the decohered density matrix in the non-decodable phase is known to admit a decomposition as a convex combination of short-range-entangled (SRE) pure states~\cite{Separability_Transitions}. In the decodable phase, no such decomposition is expected to exist, since the state retains long-range entanglement as diagnosed by the topological entanglement negativity~\cite{Fan_Bao_Altman}. The transition between the two regimes is a decoherence-induced \emph{separability transition}.


For the $\mathbb{Z}_{N>4}$ toric code, it is natural to ask how separability is to be understood within the information critical phase. We now show that in the information critical phase, the mixed state naturally decomposes into a convex combination of ``Coulombic'' pure states---states supporting gapless gauge modes that we interpret as emergent photons. This observation establishes a direct connection between information critical mixed states and conventional quantum critical pure states.

Following Refs.~\cite{METTS,TO_finite_temp,Separability_Transitions}, we use the ``minimally entangled typical thermal state'' (METTS) ansatz to write the $X$-dephased toric code density matrix, initialized in $\ket{\Psi_0}$, as the convex ensemble

\begin{align}
    \mathcal{E}(\rho_{Q}) &= \sum_{\textbf{x}} \mathcal{E}(\rho_{Q})^{1/2} | \textbf{x}\rangle \langle  \textbf{x}| \mathcal{E}(\rho_{Q})^{1/2} \nonumber \\
    &= \sum_{\textbf{x}}| \psi_{Q}({\textbf{x})}\rangle \langle \psi_{Q}(\textbf{x})|,
\end{align}
where $\ket{\textbf{x}} = \ket{\{ x_{\mu} \}}$ are the simultaneous eigenstates of all Pauli-$X$ operators. Each METTS wavefunction $\ket{\psi_{Q}(\textbf{x})}$ is obtained by ``modular evolution'' of the product state $\ket{\textbf{x}}$ by the operator $\mathcal{E}(\rho_{Q})^{1/2}$. Such a decomposition has proven to be tremendously useful in the $\mathbb{Z}_{2}$ toric code, since the states $\ket{\psi_{Q}(\textbf{x})}$ themselves undergo a phase transition from topologically ordered to topologically trivial precisely at the decodability threshold.

Since we consider only $X$-type dephasing, the plaquette constraint $B_{p}=1$ is never violated, so the decohered density matrix has support only on configurations $|\mathbf{x}\rangle$ with $B_{p}\ket{\textbf{x}} =\ket{\textbf{x}}$ for all $p$. But all such configurations can be generated from the reference state $\ket{\textbf{1}}\equiv \ket{\{ x_{\mu}=1\}}$ by acting with closed 't~Hooft loops, $\ket{\textbf{x}}  = Z_{\hat{\mathcal{C}}(\textbf{x})}\ket{\textbf{1}}$. Because these loop operators commute with $\mathcal{E}(\rho_{Q})$, the corresponding METTS states obey $\ket{\psi_{Q}(\textbf{x})} = Z_{\hat{\mathcal{C}}(\textbf{x})}\ket{\psi_{Q}(\textbf{1})}$. Thus, all $|\psi_{Q}(\mathbf{x})\rangle$ are related by finite-depth local unitaries and lie in the same quantum phase. To diagnose their topological character it therefore suffices to analyze the reference state $|\psi_{Q}(\mathbf{1})\rangle$.

Using the explicit form of the decohered density matrix, one finds
\begin{align}\label{eq:METTS_Ansatz}
    \ket{\psi_{Q}(\textbf{1})} =\sum_{\textbf{z}} \mathcal{Z}[\textbf{z}]^{1/2}\ket{\textbf{z}},
\end{align}
where $\ket{\textbf{z}} = \ket{\{ z_{\mu} \}}$ are the simultaneous eigenstates of all Pauli-$Z$ operators and
\begin{align}
    \mathcal{Z}[\textbf{z}] =  \sum_{\substack{(\nabla\cdot \textbf{k})_{v} = 0 \\ \hat{\mathcal{C}}(\textbf{k}) = 0 }} \prod_{\mu}p^x_{k_{\mu} - z_{\mu}} 
\end{align}
is again the classical partition function of a disordered clock model, now with quenched couplings $\textbf{z}$. In the low-temperature (weak-decoherence) limit, $|\psi_{Q}(\mathbf{1})\rangle \propto |\Psi_{0}\rangle$ and the state retains full topological order. In contrast, in the high-temperature (strong-decoherence) limit, $\ket{\psi_{Q}(\textbf{1})} \propto \ket{\{x_{\mu}=1 \}}$, a trivial product state.

\begin{figure}
    \centering
    \includegraphics[width=0.98\columnwidth]{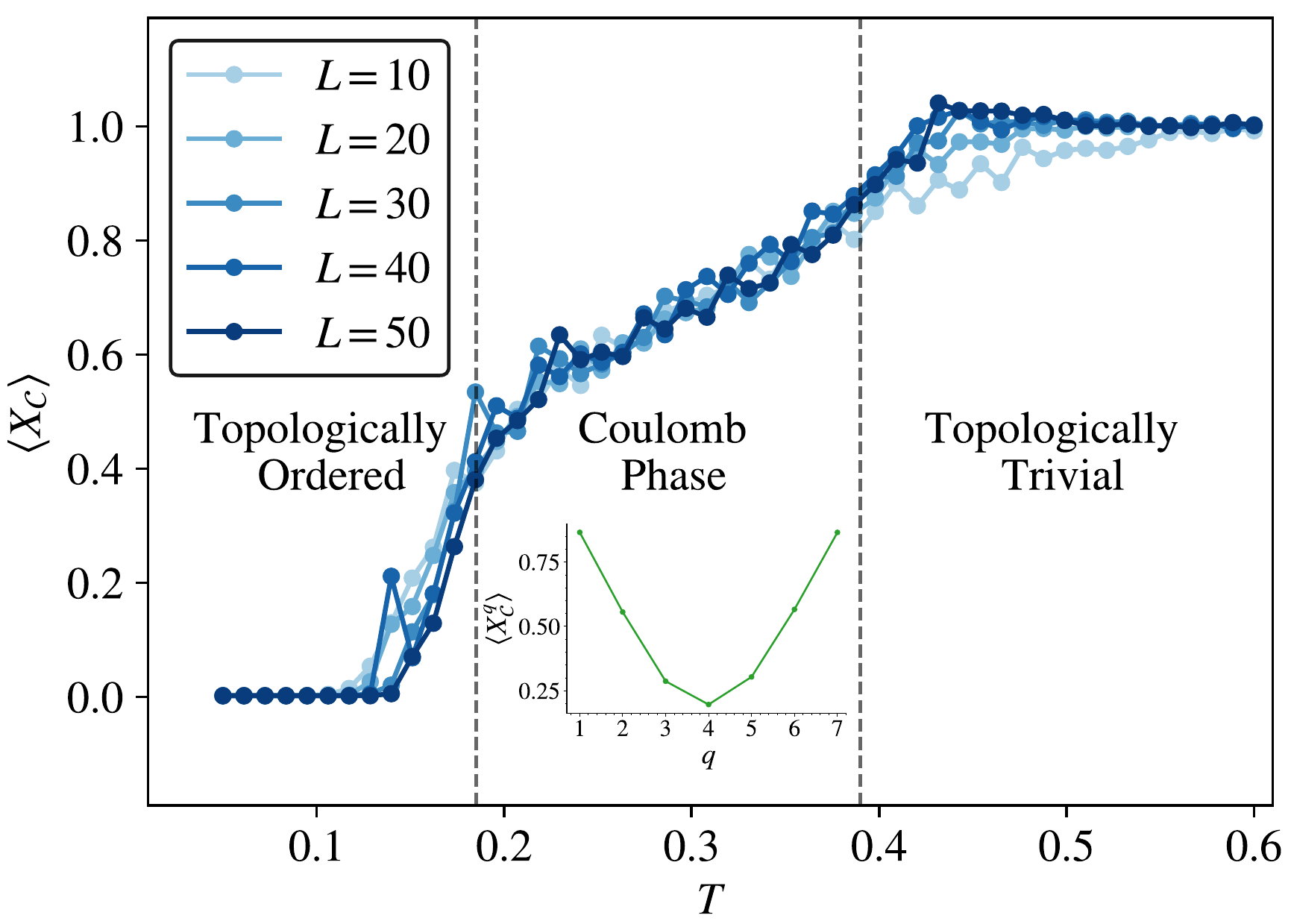}
    \caption{{\bf Wilson Loop Expectation.} We plot the expectation value of the Wilson loop operator $\langle X_{\mathcal{C}}^{q} \rangle$ (shown here for $q=1$) with respect to the METTS wavefunction $\ket{\psi_{Q}(\textbf{1})}$ (\eqnref{eq:METTS_Ansatz}) as a function of the decoherence temperature for the $\mathbb{Z}_8$ toric code. In the decodable phase, $\langle X_{\mathcal{C}}^{q} \rangle$ vanishes in the thermodynamic limit, signaling topological order, whereas in the non-decodable phase it saturates to 1 indicating topological triviality. In the information critical phase, $\langle X_{\mathcal{C}}^{q} \rangle$ approximately saturates to a constant that is roughly independent of system size and increases monotonically with decoherence strength. The inset shows the dependence of this saturation value on the loop charge $q \mod N$ within the information critical phase (at $T = 0.3$). We find that the saturation is sharply decreasing with increasing $q$ suggesting that the system can only tunnel within a restricted subset of topological sectors. Together with the power-law scaling of open Wilson strings in this phase, this behavior supports the interpretation in terms of Coulombic states with emergent gapless photons.}
    \label{fig:Wilon_Loop}
\end{figure}
A convenient diagnostic of topological order is the expectation value of a non-contractible Wilson loop $X^{q}_{\mathcal{C}}$ carrying charge $q = 1 \cdots, N-1$ with respect to the state $\ket{\psi_{Q}(\textbf{1})}$. Physically, $\langle X^{q}_{\mathcal{C}}  \rangle$ measures the amplitude for tunneling between distinct topological sectors of the $\mathbb{Z}_{N}$ toric code: it vanishes in a topologically ordered phase and saturates to an $\mathcal{O}(1)$ value in a topologically trivial phase where anyons condense. Since $X_{\mathcal{C}}\ket{\textbf{z}} = \ket{\textbf{z} + q\delta_\mathcal{C}}$, with $\delta_\mathcal{C}$ supported only on the links of the loop $\mathcal{C}$, we have
\begin{align}
    \langle X^{q}_{\mathcal{C}}\rangle  &= \sum_{\textbf{z}}\sqrt{\mathcal{Z}[\textbf{z}]\mathcal{Z}[\textbf{z}+ q\delta_\mathcal{C}]} \nonumber \\
    &= \sum_{\textbf{z}}\mathcal{Z}[\textbf{z}]\sqrt{\frac{\mathcal{Z}[\textbf{z}+ q\delta_\mathcal{C}]}{\mathcal{Z}[\textbf{z}]}} .
\end{align}
which can be interpreted as a Nishimori line expectation of a ratio of conditional winding number probabilities: $\langle X^{q}_{\mathcal{C}}\rangle = [\sqrt{P(q|\textbf{z})/P(0|\textbf{z})}]$. As such, it directly probes the disorder-renormalized spin stiffness, and can be evaluated with the same Monte Carlo techniques as in~\secref{sec:Numerical_details}. The result for the decohered $\mathbb{Z}_{8}$ toric code is shown in~\figref{fig:Wilon_Loop}.

As expected, the expectation value decays exponentially with system size in the decodable phase and saturates to an $\mathcal{O}(1)$ value in the non-decodable phase. This equates decodability with the presence of topological order in the METTS states, and non-decodability with their topological triviality. In the information critical phase, the tunneling amplitude saturates to a constant that is independent of system size, exactly as predicted by spin-wave theory. Nevertheless, this saturation value decays sharply with the loop charge. At fixed decoherence strength, the system can therefore tunnel only within a restricted subset of topological sectors since the probability of creating large-charge zero modes remains exponentially suppressed. This furnishes a simple physical picture for the fractional protection of logical information. Logical information encoded in superpositions of sectors separated by large charge differences is effectively protected, because decoherence cannot induce tunneling between them, whereas information stored in superpositions differing by small charges is efficiently destroyed. As the decoherence strength is increased, additional charge sectors are progressively unlocked, leading to a continuous erosion of the protected logical subspace.

To gain further insight into these states, we consider the expectation value of an open Wilson string $\langle X_{\gamma} \rangle$, which creates a pair of electric charges $\pm 1$ at its endpoints $\partial \gamma$. At large charge separation ($|\gamma| \gg 1$), this serves as an order parameter for detecting the condensation of electric charges in the gauge states $\ket{\psi_{Q}}$. It is plotted in~\figref{fig:Renyi1_Correlator} under a different name---the R\'enyi-1 correlator, defined in the next section. In the decodable phase $\langle X_{\gamma}\rangle\sim e^{-\alpha |\gamma|}$ decays exponentially with the charge separation, while in the non-decodable phase $\langle X_{\gamma}\rangle\sim \mathcal{O}(1)$ as $|\gamma|\rightarrow \infty$, signaling charge condensation and pointing to a ``Higgs phase.'' In the information critical phase, the open-string expectation value decays as a power law, $\langle X_{\gamma}\rangle\sim |\gamma|^{-\eta(T)}$, with an exponent $\eta(T)$ that varies continuously with decoherence strength. This signals the presence of gapless modes that we naturally interpret as emergent photons. We conclude that the states $\ket{\psi_{Q}(\textbf{x})}$ in the convex decomposition may be viewed as gapless Coulomb states.

This does not strictly rule out a decomposition of the information-critical density matrix into genuine SRE states, but we consider that scenario unlikely given the nonzero coherent information. We note that it would be interesting to directly probe the long-range entanglement structure of the information critical phase using genuine mixed state diagnostics, such as the topological entanglement negativity, and to determine whether this quantity likewise saturates to a fractional value in this regime. We leave a detailed investigation of this question for future work.

\section{Dual Strong-to-Weak SSB} \label{sec:dualSWSSB}
Our results also immediately yield the mixed-state phase diagrams of a family of decohered spin systems obtained by ``ungauging'' the $\mathbb{Z}_{N}$ toric code via Kramers-Wannier (KW) duality.
Under this duality, the flux-free sector of the $\mathbb{Z}_{N}$ toric code (corresponding to $B_p=1$ for all $p$) maps to a $\mathbb{Z}_{N}$ quantum spin model through the operator dictionary
\begin{align}\label{eq:KW_duality}
    A_v &\longleftrightarrow \bar{X}_{v} \nonumber\\
    X_{\mu = \langle v_{1} \rightarrow v_{2} \rangle} &\longleftrightarrow \bar{Z}_{v_{1}}(\bar{Z}_{v_{2}})^{\dagger}.
\end{align}
The spins now reside on the vertices, and in the $\bar{X}$ basis their eigenvalues directly encode the anyon charge at each vertex. The ``Bianchi identity'' $\prod_{v}A_{v}=1$ is promoted to a global $0$-form $\mathbb{Z}_{N}$ symmetry $\prod_{v}\bar{X}_{v}=1$, reflecting conservation of total anyon charge, while the toric code ground state $\ket{\Psi_0}$ (with $A_v=1$ for all $v$) maps to the paramagnetic product state $\otimes_{v}|\bar{X}_{v} = 1\rangle$.

We again restrict to $X$-type dephasing in the $\mathbb{Z}_{N}$ toric code. Since this noise preserves the flux-free constraint $B_p = 1$, the KW dictionary continues to hold, and the local decoherence channel on each link $\langle v\rightarrow v'\rangle$ becomes
\begin{align}\label{eq:dual_decoherence_channel}
    \mathcal{E}_{\langle v \rightarrow v' \rangle}(\rho) &= \sum_{k=0}^{N-1} p^{x}_{k}(\bar{Z}_{v}\bar{Z}^{\dagger}_{v'})^{k}\rho (\bar{Z}_{v'}\bar{Z}^{\dagger}_{v})^{k}.
\end{align}
The resulting decohered spin state can now be diagonalized in the $\bar{X}$ basis, yielding 
\begin{align}
    \mathcal{E}(\rho) = \sum_{\textbf{e}}\mathcal{Z}[\textbf{e},0]|\textbf{e}\rangle\langle\textbf{e}|,
\end{align}
with $\ket{\textbf{e}} =\otimes_{v}|\bar{X}_{v} = e_v\rangle$.

In this dual $\mathbb{Z}_{N}$-symmetric spin system, the decoherence-induced phase diagram is naturally interpreted in the language of strong-to-weak spontaneous symmetry breaking (SWSSB)~\cite{Lee_Quantum_Criticality,Olumakinde2023,Strong_Weak_SSB_Open_Systems,Strong_to_weak_Fidelity_Correlator,Strong_Weak_SSB_Renyi1_Correlator,Strong_Weak_SSB_Hydrodynamics,Strong_Weak_SSB_SYK,Strong_Weak_SSB_Wightman,Strong_Weak_SSB_1form}. Recall that a mixed state $\rho$ possesses a \emph{strong} (or exact) $G$ symmetry, with unitary representation $U:G\to\mathcal{U}(\mathcal{H})$, if
\begin{align}
    U({g})\rho = e^{i\theta_{g}}\rho.
\end{align}
for all $g\in G$. On the other hand, it is said to possess a \emph{weak} (or average) $G$ symmetry if
\begin{align}
    U({g})\rho U(g)^{\dagger} = \rho.
\end{align}
Operationally, strong symmetry means every pure state in an ensemble decomposition of $\rho$ carries the same symmetry charge, whereas weak symmetry allows the ensemble to contain states with different (but individually well-defined) charges. Physically, which symmetry is realized depends on whether the system effectively exchanges $G$-charged particles with the environment: strong symmetry arises in a ``canonical'' setting where such exchange is forbidden, while weak symmetry is characteristic of a ``grand canonical'' setting where charge exchange is allowed.

The distinction between strong and weak symmetries also extends to quantum channels. Let $\mathcal{E}(\cdot) = \sum_{i} \mathcal{K}_{i}(\cdot)\mathcal{K}_{i}^{\dagger}$ be a generic channel with Kraus operators $\mathcal{K}_{i}$. The channel has a weak $G$ symmetry if it is covariant for all $\rho$,
\begin{align}
    \mathcal{E}\bigg(U_{g}\rho  U_{g}^{\dagger} \bigg) = U_{g}\mathcal{E}(\rho)  U_{g}^{\dagger},
\end{align}
and a strong $G$ symmetry if each Kraus operator commutes with the symmetry,
\begin{align}
    [\mathcal{K}_{i}, U_{g}] = 0.   
\end{align}
\begin{figure}
    \centering
    \includegraphics[width=0.98\columnwidth]{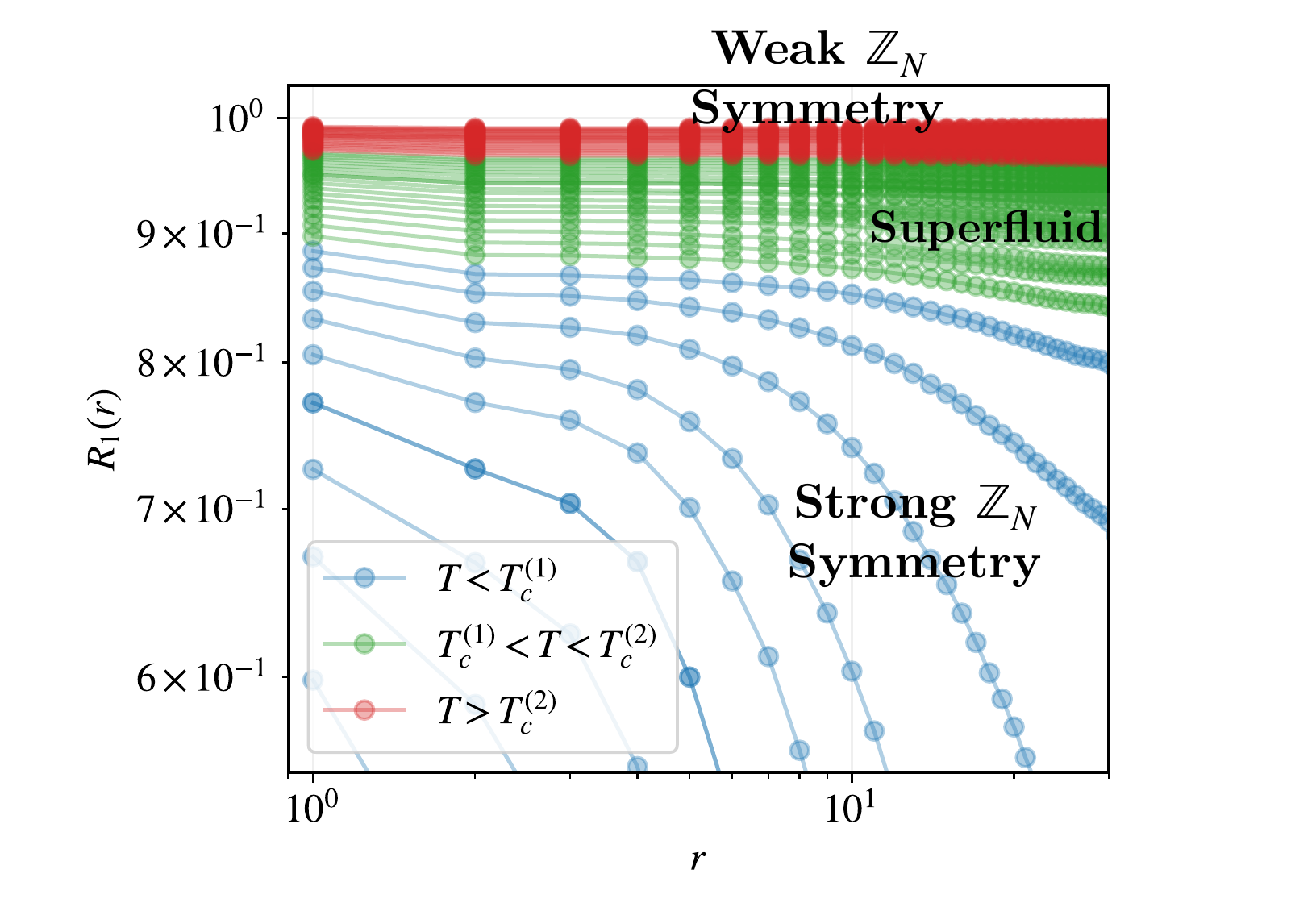}
    \caption{{\bf R\'enyi-1 Correlator.} We plot the R\'enyi-1 correlator in the $\mathbb{Z}_8$ symmetric model, which is the Kramers-Wannier dual of the decohered $\mathbb{Z}_{8}$ toric code~\cite{Lee_Quantum_Criticality} for a range of decoherence strengths, where $r$ denotes the separation between the dual spins. We find that for $T<T^{(1)}_{c}$, corresponding to the decodable phase of the toric code, $R_{1}(r) \sim e^{-\alpha r}$ indicating the persistence of strong $\mathbb{Z}_{N}$ symmetry. For $T>T^{(2)}_{c}$, corresponding to the non-decodable phase, $R_{1}(r) \sim \mathcal{O}(1)$ as $r \rightarrow \infty$ signaling SWSSB. For $T^{(1)}_{c}\leq T\leq T^{(2)}_{c}$, we find that $R_{1}(r)\sim r^{-\eta(T)}$ exhibits an approximate power-law scaling. This points to a superfluid phase with quasi-long-range order.}
    \label{fig:Renyi1_Correlator}
\end{figure}
SWSSB occurs when a strongly symmetric channel spontaneously breaks the strong symmetry of an initial state down to a weak symmetry. A particularly transparent way to understand it is to recast it as an ordinary SSB transition in a doubled Hilbert space. Following Ref.~\cite{Strong_Weak_SSB_Renyi1_Correlator}, we consider the \emph{canonical purification} of $\rho$---the Choi-Jamio\l{}kowski vectorization~\cite{CHOI1975,JAMIOLKOWSKI1972} of the operator $\sqrt{\rho}:\mathcal{H}\to\mathcal{H}$---which yields the state $\Vert\sqrt{\rho}\rangle\!\rangle\in\mathcal{H}\otimes\mathcal{H}^*$,
given by
\begin{align}
    \Vert\sqrt{\rho}\rAngle = (\sqrt{\rho}\otimes 1)\sum_{\textbf{s}}|\textbf{s}\rangle_{L} |\textbf{s}\rangle_{R}.
\end{align}
Here $L$ is the original Hilbert space with orthonormal basis $\ket{\textbf{s}}$ and $R$ its charge-conjugate counterpart. By construction $\Vert \sqrt{\rho}\rAngle$ purifies $\rho$,
\begin{align}
    \rho = \Tr_{R} \Vert\sqrt{\rho}\rAngle \lAngle \sqrt{\rho}\Vert.
\end{align}
If $\rho$ has strong $G$ symmetry, $\Vert\sqrt{\rho}\rangle\!\rangle$ is invariant under $G^{L}\times G^{R}$, an independent $G$ action on each copy of the Hilbert space. By contrast, if $\rho$ has only weak $G$ symmetry, the purification is invariant solely under the diagonal (``vector'') subgroup $G^{V}\subset G^{L}\times G^{R}$. SWSSB in the mixed state is thus a conventional SSB transition in the canonical purification, in which $G^{L}\times G^{R}$ spontaneously breaks down to $G^{V}$. The off-diagonal (``axial'') subgroup $G^{A}$ is spontaneously broken.

To sharply diagnose such a transition, we look for an operator in the doubled Hilbert space that is charged under $G^{L}\times G^{R}$ but neutral under $G^{V}$, and ask whether it develops long-range order. Let $\mathcal{O}^{L}_{x}(\mathcal{O}^{R}_{x})^{*}$ be such an operator. A natural diagnostic of SWSSB is then the \emph{R\'enyi-1 correlator}~\cite{Strong_Weak_SSB_Renyi1_Correlator}
\begin{align}
    R_{1}(x-y) &= \lAngle \sqrt{\rho}|\mathcal{O}^{L}_{x}(\mathcal{O}^{R}_{x})^{*}(\mathcal{O}^{L}_{y}(\mathcal{O}^{R}_{y})^{*})^{\dagger} | \sqrt{\rho} \rAngle \nonumber \\
    &= \Tr{\sqrt{\rho} \mathcal{O}_{x}(\mathcal{O}_{y})^{\dagger}\sqrt{\rho} \mathcal{O}_{y}(\mathcal{O}_{x})^{\dagger}}.
\end{align}
As $r=|x-y|\to\infty$, a nonvanishing asymptote $R_{1}(r)\sim \mathcal{O}(1)$ signals the spontaneous breaking of $G^{L}\times G^{R}$ down to $G^{V}$. The correlator $R_{1}$ is especially convenient because it admits a direct interpretation as an ordinary SSB correlator in the doubled Hilbert space, analogous to the more familiar R\'enyi-2 correlator~\cite{Lee_Quantum_Criticality}. It also inherits the stability property proved for the related \emph{fidelity correlator}~\cite{Strong_to_weak_Fidelity_Correlator}: mixed states on opposite sides of an SWSSB transition cannot be two-way connected by finite-depth channels that preserve the strong symmetry. For stabilizer states subject to Pauli decoherence channels, the R\'enyi-1 and fidelity correlators are in fact exactly equal~\cite{Strong_Weak_SSB_Renyi1_Correlator}.

Returning to the KW dual spin model, we note that the decoherence channel~\eqnref{eq:dual_decoherence_channel} preserves the strong $\mathbb{Z}_{N}$ symmetry generated by $\prod_{v}\bar{X}_{v}$, and the canonically purified decohered density matrix is simply
\begin{align}
    \Vert \sqrt{\rho}\rAngle = \sum_{\substack{\textbf{e}}} \bigg[\sqrt{\mathcal{Z}[\textbf{e},0]} \bigg] \ket{\textbf{e}}_{L}\ket{-\textbf{e}}_{R}.
\end{align}
To test for SWSSB, we compute the R\'enyi-1 correlator of the local operator $\bar{Z}^{L}_{x}(\bar{Z}^{R}_{x})^{*}$, which is charged under $\mathbb{Z}^{L}_{N}\times \mathbb{Z}^{R}_{N}$ but neutral under the vector subgroup $\mathbb{Z}^{V}_{N}$. Consequently, long-range order in this correlator signals SWSSB of the decohered density matrix. Explicitly, we find that the R\'enyi-1 correlator takes the following form
\begin{align}\label{eq:defect_defect_correlator}
    R_{1}(x-y) = \sum_{\textbf{e}}\mathcal{Z}[\textbf{e},0]\sqrt{\frac{\mathcal{Z}[\textbf{e} + \textbf{e}_{xy},0]}{\mathcal{Z}[\textbf{e},0]}},
\end{align}
where $\textbf{e}_{xy}$ denotes charge insertions of magnitude $\pm1$ at $x$ and $y$. This is just a disorder-averaged defect-defect correlator of the classical spin model along the Nishimori line, which we evaluate numerically with the techniques of~\secref{sec:Numerical_details}. We depict the resulting plot for the $\mathbb{Z}_{8}$ toric code in~\figref{fig:Renyi1_Correlator}.

For $N>4$, the defect-defect correlator $R_{1}(r)$ exhibits three distinct behaviors across the phase diagram. At low $T$, when the toric code is in the decodable phase, $R_{1}(r) \sim e^{-\alpha r}$ decays exponentially, indicating no long-range order and the persistence of the strong $\mathbb{Z}_{N}^{L}\times \mathbb{Z}_{N}^{R}$ symmetry. Anyons remain closely confined together in this phase. At high $T$, when the toric code is \emph{non-decodable}, the correlator saturates to a nonzero constant, $R_{1}(r)\sim \mathcal{O}(1)$, signaling the breaking of $\mathbb{Z}_{N}^{L}\times \mathbb{Z}_{N}^{R}$ down to the diagonal $\mathbb{Z}_{N}^{V}$. Although the net anyon charge always remains fixed (at zero), anyon pairs proliferate in this phase. 

In the intermediate information critical phase, the R\'enyi-1 correlator instead follows a power law $R_{1}(r) \sim r^{-\eta(T)}$, with an exponent $\eta(T)$ that varies with the decoherence temperature. This behavior again arises because in this regime, discrete $\mathbb{Z}_{N}$ locking terms become RG-irrelevant, leading to an effective enhancement of the symmetry to a continuous $U(1)$ group at long distances. Consequently, the infrared physics of the decohered $\mathbb{Z}_{N}$ spin model here closely parallels that of a decohered $U(1)$ XY model. This effective symmetry enhancement of the broken (axial) $\mathbb{Z}^{A}_{N}$ symmetry to a $U(1)^{A}$ symmetry places the system in a novel \emph{superfluid phase} with quasi-long-range order. The gapless modes responsible for the power-law decay in Eq.~\eqref{eq:defect_defect_correlator} couple both copies of the Hilbert space, and may be viewed as collective excitations of the system and the environment together. The full SWSSB pattern for the KW dual of the decohered $\mathbb{Z}_{N>4}$ toric code, illustrated in \figref{fig:Summary_Figure}(b), is therefore
\begin{align}
    \textrm{Fully symmetric} \longrightarrow U(1)^{A} \textrm{ QLRO}  \longrightarrow \mathbb{Z}_{N}^{A} \textrm{ SSB}. \nonumber
\end{align}
Since the R\'enyi-1 correlator lower-bounds the CMI~\cite{Strong_to_weak_Fidelity_Correlator, Strong_Weak_SSB_Renyi1_Correlator}, this power-law decay immediately forces $\xi_M$ to diverge in the dual model as well. Finally, note that the R\'enyi-1 correlator considered here is exactly equal to the open Wilson-string expectation value $\langle X_{\gamma} \rangle$ evaluated in the METTS wavefunction $\ket{\psi_{Q}(\textbf{1})}$ of~\secref{sec:separability}. This correspondence provides a transparent link between the SWSSB pattern of the dual spin model and the topological properties encoded in the METTS wavefunctions.


\section{Maximum Likelihood Decoding} \label{sec:decoding}

We now address decoding of the noise-corrupted $\mathbb{Z}_{N}$ toric code. We implement maximum-likelihood decoding using a Monte Carlo worm algorithm interspersed with parallel-tempering swaps. The coherent-information analysis of the previous section sets a fundamental, information-theoretic bound on the performance of any decoder.
\begin{figure}
    \includegraphics[width=0.99\columnwidth]{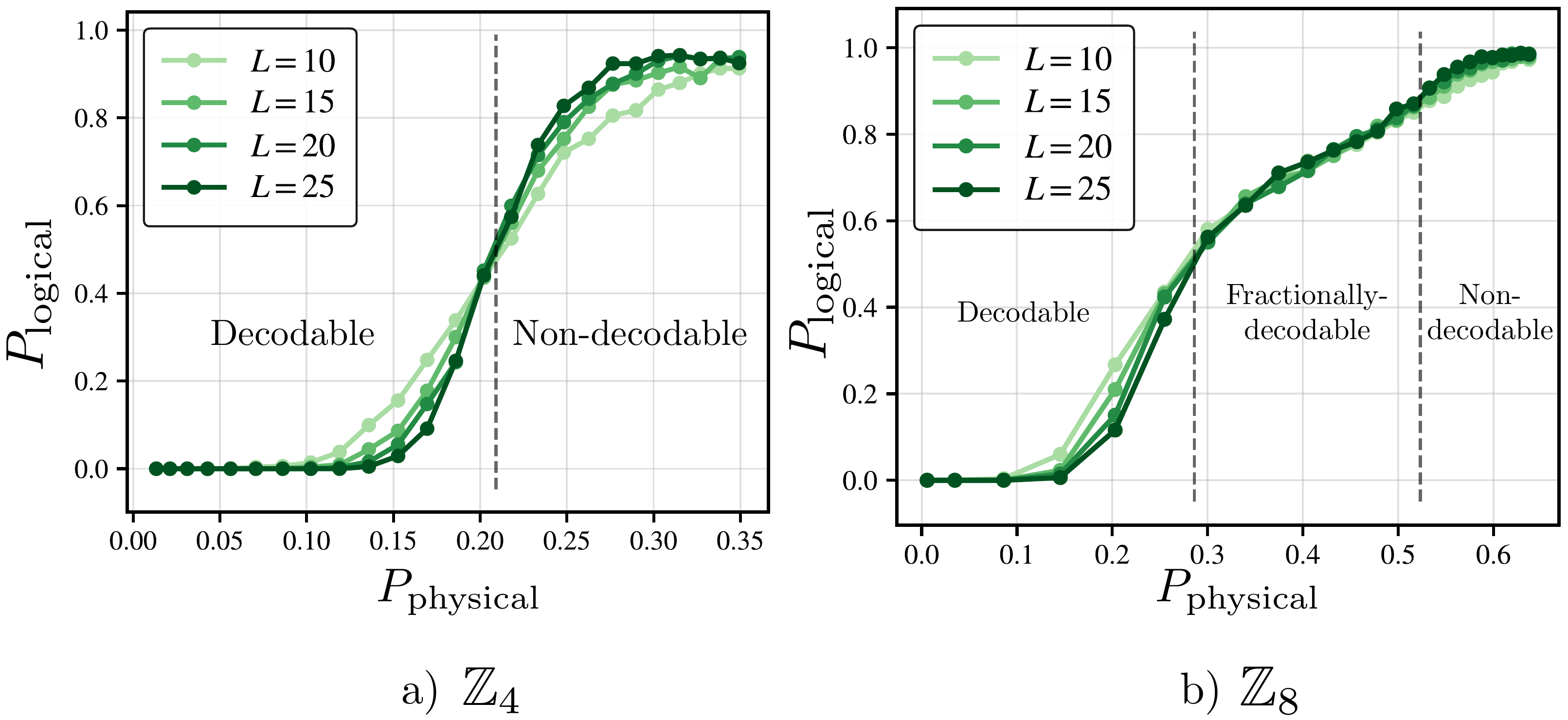}
    \caption{\label{fig:Decoding_numerics}  {\bf Decoding.} Performance of the proposed maximum-likelihood decoder, which combines worm Monte Carlo sampling with parallel tempering. We plot the logical error rate $P_{\text{logical}}$ as a function of the physical error rate for the \textbf{a)} $\mathbb{Z}_{4}$ and \textbf{b)} $\mathbb{Z}_{8}$ toric codes, averaged over 1000 error realizations. For $\mathbb{Z}_{4}$, the curves cross at a single threshold that coincides with the decodable--non-decodable transition of the decohered toric code, confirming the optimality of the decoder. For $\mathbb{Z}_{8}$, the code enters an intervening fractionally decodable phase where $P_{\text{logical}}$ converges in the $L\to\infty$ limit to a finite, fractional value that grows continuously with the decoherence strength. The two thresholds agree with the critical points identified from the coherent information. This provides the operational signature of a fractional quantum memory.
    } 
\end{figure}

Efficient decoding of the decohered toric code has a rich history. Decoding is essentially a classical post-processing task: given the syndrome-measurement outcomes, one seeks a combination of $X$ and $Z$ string operators that cancels the syndromes and returns the system to the codespace without incurring a logical error. More formally, suppose the toric code is subject only to $X$-type decoherence with an arbitrary parameter set $\{ p^{x}_{k}\}$. This produces an incoherent mixture of $e$ anyons. Syndrome measurements then read off the anyon charges $\mathbf{e} = \{ e_{v}\}$ at each vertex, and the task reduces to finding a flow $\textbf{k}$ that solves the divergence constraint $(\nabla \cdot \textbf{k})_{v} = e_{v}$, so that the syndromes can be cancelled by applying $(X^{\textbf{k}})^{\dagger}$. There exist several different flow configurations which solve this constraint, all related to one another through the addition of divergence free flows. The relative probabilities of these distinct solutions depend on the noise parameters $\{ p^{x}_{k}\}$. Thus, decoding is not merely about feasibility but is rather an optimization problem over equivalence classes of flows. 

Within this framework one distinguishes \emph{hard-decision} from \emph{soft-decision} strategies. A hard-decision decoder selects a single flow configuration to cancel the measured syndromes, succeeding as long as the chosen configuration introduces no spurious non-contractible loops that would implement a logical error. For the $\mathbb{Z}_{2}$ toric code, the Minimum Weight Perfect Matching (MWPM) algorithm~\cite{Dennis_2002_Error_Threshold}---efficiently implemented via Edmonds' blossom algorithm~\cite{Blossoms_algo}---is a powerful hard-decision decoder with a high threshold. Hard-decision RG decoders are another prominent example~\cite{RG_decoder_hard_decisions,RG_Decoders_Zn_hard_decisions}. Soft-decision decoders, by contrast, are probabilistic: they assign likelihoods to different flow configurations and return a distribution over all logical sectors. Examples include maximum-likelihood decoding and the soft-decision RG decoders in Refs.~\cite{RG_Decoder_soft_decisions1,RG_decoder_soft_decisions2,RG_Decoders_Zn_soft_decisions, RG_Decoders_Zn_Watson}. Broadly speaking, hard-decision decoders tend to be computationally cheaper and faster, whereas soft-decision decoders tend to achieve better thresholds.

The optimal maximum-likelihood decoder does not select a single error but rather the most probable logical class. Two flows with the same divergence and the same winding, $\hat{\mathcal{C}}(\textbf{k})=\textbf{a}$, differ only by stabilizers and implement the same recovery. We say they belong to the same logical coset. The probability of a coset is obtained by summing the noise weight over all flows it contains, which is exactly the partition function introduced in \secref{sec:diag},
\begin{align}\label{eq:ML_coset_prob}
    P(\textbf{a}\,|\,\textbf{e}) = \frac{\mathcal{Z}[\textbf{e},\textbf{a}]}{\sum_{\textbf{a}'}\mathcal{Z}[\textbf{e},\textbf{a}']}, \qquad
    \mathcal{Z}[\textbf{e},\textbf{a}] = \!\!\sum_{\substack{(\nabla\cdot\textbf{k})_v = e_v \\ \hat{\mathcal{C}}(\textbf{k})=\textbf{a}}}\!\!\prod_{\mu}p^{x}_{k_{\mu}}.
\end{align}
Maximum-likelihood decoding then returns the most probable class $\textbf{a}_\ast = \arg\max_{\textbf{a}}\mathcal{Z}[\textbf{e},\textbf{a}]$. The difficulty here is that each $\mathcal{Z}[\textbf{e},\textbf{a}]$ is a sum over an exponentially large set of stabilizer-equivalent flows. Evaluating such partition functions exactly is in general $\#$P-hard. We therefore estimate the coset probabilities $P(\textbf{a}\,|\,\textbf{e})$ by Monte Carlo rather than computing them directly.

Concretely, given the measured syndromes $\{e_{v}\}$ we first construct a deterministic reference flow $\textbf{k}^{0}$ solving $(\nabla\cdot\textbf{k}^{0})_v = e_v$. This step is purely geometric and ignores the noise probabilities; its only role is to fix one representative flow, after which every admissible configuration is written as $\textbf{k} = \textbf{k}^{0} + \textbf{l}$ with $\textbf{l}$ divergence-free. We sample the fluctuation $\textbf{l}$ using the worm algorithm, whose updates grow closed loops of charge that can wind around the non-contractible cycles of the torus. This update can therefore move the configuration both within and between homology sectors. The Boltzmann weights $\prod_{\mu}p^{x}_{k^{0}_{\mu}+l_{\mu}}$ are set by the physical noise parameters. Histogramming the winding numbers $\hat{\mathcal{C}}(\textbf{l})$ visited during the sampling directly estimates the conditional distribution $P(\textbf{a}\,|\,\textbf{e})$ in~\eqnref{eq:ML_coset_prob}.

Unfortunately, this sampling, by itself, is hampered by severe critical slowing down in the QLRO phase. There the worm trajectories have a finite probability of growing exponentially long before closing, so the autocorrelation time of the Monte Carlo---and in particular the time to tunnel between distinct homology classes---grows sharply. While the chain remains formally ergodic, on any accessible timescale it effectively stalls within a single topological sector, precisely where we need to resolve the competition between sectors. We overcome this with \emph{parallel tempering}: we evolve a ladder of replicas with tempered weights $\big(p^{x}_{k}\big)^{\lambda}$ for $\lambda\in[0,1]$, interpolating between the physical noise at $\lambda=1$ and the uniform, maximally disordered distribution at $\lambda=0$, where the sectors are sampled freely. Adjacent replicas are periodically exchanged via the standard Metropolis criterion, so that swaps with the low-$\lambda$ replicas inject topological tunneling events into the physical replica, dramatically reducing the equilibration time. The coset histogram is accumulated only from the physical replica at $\lambda=1$, and decoding is completed by selecting the most probable sector $\textbf{a}_\ast$ and applying the correction $\big(X^{\textbf{a}_\ast}\big)^{\dagger}\big(X^{\textbf{k}^{0}}\big)^{\dagger}$. A logical error is recorded whenever $\textbf{a}_\ast$ differs from the true logical class of the sampled error.

The performance of this decoder is shown in~\figref{fig:Decoding_numerics}, where we plot the logical error probability $P_{\mathrm{logical}}$ against the physical error rate $p_{\mathrm{physical}} = 1-p^{x}_{0}(T^x)$. For the $\mathbb{Z}_{4}$ code, the curves for different system sizes cross sharply at a single threshold that agrees, to good accuracy, with the decodable--non-decodable critical point identified from the coherent information, confirming the optimality of the decoder. For the $\mathbb{Z}_{8}$ code, the data reveal three regimes: below the first threshold the code is decodable and $P_{\mathrm{logical}}\to 0$ as $L\to\infty$, while above the second it is non-decodable and $P_{\mathrm{logical}}$ approaches $1-1/N^{2}$, the random-guessing value. In the intervening information critical phase, $P_{\mathrm{logical}}$ instead saturates in the thermodynamic limit to a fractional value that varies continuously with the decoherence strength. This is the operational signature of a fractional quantum memory, in which some logical information is perfectly protected while the rest is irretrievably lost. 
The decoder still reliably resolves the logical sectors separated by large charge but is reduced to guessing among those separated by small charge, which decoherence has effectively mixed together. The code thus behaves as though it retained a reduced effective logical dimension $N_{\rm eff}<N$ that shrinks continuously with increasing decoherence strength.

\section{Discussion}
In this work we have enlarged the landscape of mixed-state phases by introducing the \emph{information critical phase}: an extended region of phase space in which the Markov length diverges while the ordinary correlation length stays finite. We showed that such a phase is realized in decohered $\mathbb{Z}_{N}$ toric codes for $N>4$. In the information critical phase, the code functions as a fractional quantum memory that preserves a finite fraction of its logical information. We further argued that the decohered state in this phase admits a decomposition into a convex ensemble of Coulombic pure states. In the ungauged, Kramers-Wannier dual of the $\mathbb{Z}_{N}$ toric code, the information critical phase corresponds to a superfluid-like phase whose gapless modes arise from collective fluctuations of the system and the environment together. From the perspective of strong-to-weak SSB, the information critical phase emerges through an effective enhancement of the broken axial $\mathbb{Z}_{N}$ symmetry to a continuous $U(1)$ symmetry. Finally, we introduced a maximum-likelihood decoding algorithm for the corrupted $\mathbb{Z}_{N}$ toric code, based on worm Monte Carlo sampling combined with parallel tempering. We found that in the information critical phase the logical error rate converges to a fractional value in the thermodynamic limit, reinforcing the picture of the code as a fractional memory with a reduced logical subspace.

Several directions for future work follow naturally from our results. First, it would be interesting to clarify how one should \emph{characterize} an information critical phase. Conventional criticality admits an effective CFT description, organized by universal data such as scaling dimensions and operator-product coefficients. By contrast, in the family of examples studied here, criticality is realized through a \emph{disordered ensemble} of critical states, so that different observables can be governed by different effective scaling exponents. Indeed, disordered critical systems are known to host an infinite family of scaling dimensions, depending on both the observable and the disorder moment probed~\cite{LUDWIG1987687, LUDWIG1990639}. Developing an analogous set of universal descriptors for information criticality, from both information-theoretic and field-theoretic viewpoints, remains an important open problem. Relatedly, it would be valuable to analyze the information critical phase within the recently proposed mixed-state bootstrap framework~\cite{Lee_Bootstrap_Mixed_State}---for instance, to sharpen the characterization of the scale-invariant conditional mutual information in this phase.


Furthermore, our analysis was restricted to charge-conjugation-symmetric dephasing channels ($p_{k}= p_{N-k}$), and it is natural to ask what new mixed-state phases arise once this symmetry is broken. In the corresponding statistical-mechanical description, we obtain a \emph{chiral} clock model~\cite{Subir_chiral,Samajdar_2018}, whose Hamiltonian carries a sign problem and is therefore much harder to analyze. Exploring this more generic setting could reveal a richer family of mixed-state phases and contribute toward a more general classification of decoherence-induced phases.

Finally, our study of the dual $\mathbb{Z}_N$ SWSSB transition indicates that, under a strongly $\mathbb{Z}_N$ symmetric Lindbladian evolution, there exists a time range where the system may exhibit \emph{emergent} $U(1)$ symmetry. As the presence of strong $U(1)$ symmetry in the Lindbladian dynamics is related to the presence of diffusive behaviors~\cite{Olumakinde2023, Sanjay2024}, it would be interesting to study whether the presence of emergent continuous symmetry can give rise to novel diffusive behaviors in intermediate timescales.

\vspace{5pt}

\emph{Notes added}: While preparing this manuscript, we became aware of an independent work on decohered $U(1)$ symmetry enriched topological codes, which implements minimum integer flow alongside worm decoders~\cite{Tekmin2025}. 
We also point readers to Ref.~\cite{Optimal_decoding_worm} for more in-depth discussion of worm decoders.

\acknowledgments

We thank Tai-Hsuan Yang, Andreas Ludwig, Ehud Altman, and Yuta Hirasaki for several fruitful discussions. In particular, we thank Zack Weinstein for directing us to the Worm algorithm that significantly expedited the numerical simulation. 
The work is supported by the faculty startup grant at the University of Illinois, Urbana-Champaign and the IBM-Illinois Discovery Accelerator Institute. Additionally, JYL and AV acknowledge support from KIAS through the Quantum Universe Center scholar program.

\bibliography{refs}
\appendix 
\section{Diagonalization of decohered density matrices}\label{app:diagonalizeQ}
In this appendix we fill in the details of the analytic diagonalization of the decohered density matrices.
\subsection{Diagonalization of $\mathcal{E}(\rho_{Q})$}
Denote
\begin{align}
    \rho^{\textbf{a},\textbf{b}}_{\textbf{e};\textbf{m}} = |\textbf{e},\textbf{m},\textbf{a}\rangle\langle\textbf{e},\textbf{m},\textbf{b} |.
\end{align}
Since $\mathcal{E}(\rho_{Q})$ commutes with all stabilizer operators, we may expand it as
\begin{align}\label{eq:Diagonal_Density_Matrix}
    \mathcal{E}(\rho_{Q}) = \sum_{\substack{\textbf{e};\textbf{m} \\ \textbf{a},\textbf{b}}} \Tr{\mathcal{E}(\rho_{Q})\rho^{\textbf{b},\textbf{a}}_{\textbf{e};\textbf{m}}} \rho^{\textbf{a},\textbf{b}}_{\textbf{e};\textbf{m}},
\end{align}
so diagonalization reduces to evaluating the trace $\Tr{\mathcal{E}(\rho_{Q})\rho^{\textbf{a},\textbf{b}}_{\textbf{e};\textbf{m}}}$. Writing
\begin{align}
     \mathcal{E}(\rho_{Q}) &=  \bigg( \prod_\mu \mathcal{E}^{x}_{\mu}\circ  \prod_\mu \mathcal{E}^{z}_{\mu}\bigg)(\rho_{Q}) \\
     &= \sum_{\{ \textbf{k}^{x}, \textbf{k}^{z}\}} \bigg(\prod_{\mu}p^x_{k^{x}_{\mu}} p^z_{k^{z}_{\mu}}\bigg) \times \nonumber \\
     &\quad  Z^{\textbf{k}^{z}} X^{\textbf{k}^{x}}\rho_{Q} (X^{\dagger})^{\textbf{k}^{x}}(Z^{\dagger})^{\textbf{k}^{z}},
\end{align}
where the sum runs over all configurations $\{ \textbf{k}^{x}, \textbf{k}^{z}\}$ and
\begin{align}
    X^{\textbf{k}^{x}} &= \prod_{\mu } X^{k_{\mu}^{x}}_{\mu}   \quad,  \quad Z^{\textbf{k}^{z}} = \prod_{\mu } Z^{k_{\mu}^{z}}_{\mu}.
\end{align}
Substituting $\rho_{Q} =\frac{1}{N^{2}} \sum_{\textbf{c}}X^{\textbf{c}}|{\Psi_{0}}\rangle \langle \Psi_{0}| (X^{\textbf{c}})^{\dagger}$ into the trace yields
\begin{align}
    \Tr{\mathcal{E}(\rho_{Q})\rho^{\textbf{a},\textbf{b}}_{\textbf{e};\textbf{m}}}  &= \frac{1}{N^{2}}\sum_{\textbf{c}}\sum_{\{ \textbf{k}^{x}, \textbf{k}^{z}\}} \bigg(\prod_{\mu}p^x_{k^{x}_{\mu}} p^z_{k^{z}_{\mu}} \bigg) \nonumber \\
     \times \delta_{\textbf{a}\textbf{b}}& |\mel{\textbf{e};\textbf{m};  \textbf{a}}{ Z^{\textbf{k}^{z}} X^{\textbf{k}^{x}} X^{\textbf{c}}}{\Psi_0}|^{2}.
\end{align}
The matrix element $|\mel{\textbf{e};\textbf{m};  \textbf{a}}{Z^{\textbf{k}^{z}} X^{\textbf{k}^{x}}X^{\textbf{c}}}{\Psi_0}|$ vanishes unless the open Wilson strings in $X^{\textbf{k}^{x}}$ create charges ${e}_v$ at their ends, $(\nabla\cdot \textbf{k}^x)_{v} = e_v \mod N$, and the open 't Hooft strings in $Z^{\textbf{k}^{z}}$ create fluxes ${m}_p$ at their ends, $(\nabla\times \textbf{k}^m)_{p} = m_p \mod N$. Furthermore, $X^{\textbf{k}^{x}}X^{\textbf{c}}\ket{\Psi_0}$ must be an eigenstate of $Z_{\hat{\mathcal{C}}_{2}}$ and $Z_{\hat{\mathcal{C}}_{1}}$ with eigenvalues $a_{1}$ and $a_{2}$, i.e. $\hat{\mathcal{C}}(\textbf{k}^{x})=\textbf{a}-\textbf{c}$. When these conditions hold the matrix element has unit modulus, and otherwise it vanishes. The trace therefore reduces to
\begin{align}
    \Tr{\mathcal{E}(\rho_{Q})\rho^{\textbf{a},\textbf{b}}_{\textbf{e};\textbf{m}}}  &= \frac{1}{N^{2}}\sum_{\textbf{c}}\sum_{\substack{\textbf{k}^{z} \\ (\nabla\times \textbf{k}^z)_{p} = m_p}} \prod_{\mu} p^z_{k^{z}_{\mu}}  \nonumber \\
    &\sum_{\substack{\textbf{k}^{x} \\ (\nabla\cdot \textbf{k}^x)_{v} = e_v \ , \  \hat{\mathcal{C}}(\textbf{k}^{x}) = \textbf{a}-\textbf{c} }} \prod_{\mu}p^x_{k^{x}_{\mu}}.
\end{align}
The sum over $\mathbf{c}$ effectively removes the constraint $\hat{\mathcal{C}}(\textbf{k}^{x}) = \textbf{a}-\textbf{c}$. Defining
\begin{align}
    \mathcal{Z}[\textbf{e},\textbf{a}] &= \sum_{\substack{\textbf{k}^{x} \\ (\nabla\cdot \textbf{k}^x)_{v} = e_v \ ,\ \hat{\mathcal{C}}(\textbf{k}^{x}) = \textbf{a} }} \prod_{\mu}p^x_{k^{x}_{\mu}}, \\
    \mathcal{Z}[\textbf{m},\textbf{b}] &= \sum_{\substack{\textbf{k}^{z} \\ (\nabla\times \textbf{k}^z)_{p} = m_p \ ,\ \mathcal{C}(\textbf{k}^{z}) = \textbf{b} }} \prod_{\mu}p^z_{k^{z}_{\mu}},
\end{align}
the trace becomes
\begin{align}
     \Tr{\mathcal{E}(\rho_{Q})\rho^{\textbf{a},\textbf{b}}_{\textbf{e};\textbf{m}}} = \frac{1}{N^{2}} \mathcal{Z}[\textbf{e}] \mathcal{Z}[\textbf{m}],
\end{align}
with $\mathcal{Z}[\textbf{e}] =\sum_{\textbf{a}}\mathcal{Z}[\textbf{e},\textbf{a}]$ and $\mathcal{Z}[\textbf{m}] =\sum_{\textbf{b}}\mathcal{Z}[\textbf{m},\textbf{b}]$. We are thus left with
\begin{align}
    \mathcal{E}(\rho_{Q}) = \sum_{\substack{\textbf{e};\textbf{m} \\ \textbf{a}}} \frac{1}{N^{2}} \mathcal{Z}[\textbf{e}] \mathcal{Z}[\textbf{m}]  \rho^{\textbf{a},\textbf{a}}_{\textbf{e};\textbf{m}}.
\end{align}
\subsection{Diagonalization of $\mathcal{E}(\rho_{QR})$}\label{app:diagonalizeQR}
To compute the coherent information we also diagonalize $\mathcal{E}(\rho_{QR}) = \mathcal{E}(|\Psi\rangle_{QR} \langle\Psi |_{QR})$. Recall that
\begin{align}
    \rho_{QR} =\frac{1}{N^{2}} \sum_{\textbf{a}\textbf{a}'}X^{\textbf{a}}|{\Psi_{0}}\rangle \langle \Psi_{0}| (X^{\textbf{a}'})^{\dagger} \otimes |\textbf{a}\rangle \langle \textbf{a}'|.
\end{align}
Since the channel acts only on the toric code degrees of freedom,
\begin{align}
    \mathcal{E}(\rho_{QR})  =\frac{1}{N^{2}} \sum_{\textbf{a}\textbf{a}'}\mathcal{E}\bigg(X^{\textbf{a}}|{\Psi_{0}}\rangle \langle \Psi_{0}| (X^{\textbf{a}'})^{\dagger}\bigg) \otimes |\textbf{a}\rangle \langle \textbf{a}'|.
\end{align}
Each $\mathcal{E}\big(X^{\textbf{a}}|{\Psi_{0}}\rangle \langle \Psi_{0}| (X^{\textbf{a}'})^{\dagger}\big)$ commutes with all stabilizers, so we expand it in the Hamiltonian basis,
\begin{align}
    \mathcal{E}&(X_{\textbf{a}}|{\Psi_{0}}\rangle \langle \Psi_{0}| X_{\textbf{a}'}^{\dagger}) =  \nonumber \\
    &\sum_{\substack{\textbf{e};\textbf{m} \\ \textbf{b},\textbf{c}}} \Tr{\mathcal{E}(X_{\textbf{a}}|{\Psi_{0}}\rangle \langle \Psi_{0}| X_{\textbf{a}'}^{\dagger})\rho^{\textbf{b},\textbf{c}}_{\textbf{e};\textbf{m}}}\rho^{\textbf{c},\textbf{b}}_{\textbf{e};\textbf{m}}.
\end{align}
Substituting the channel gives
\begin{align}
    \Tr{\mathcal{E}(X_{\textbf{a}}|{\Psi_{0}}\rangle \langle \Psi_{0}| X_{\textbf{a}'}^{\dagger})\rho^{\textbf{b},\textbf{c}}_{\textbf{e};\textbf{m}}}  = \sum_{\{ \textbf{k}^{x}, \textbf{k}^{z}\}} \bigg(\prod_{\mu}&p^{x}_{k^{x}_{\mu}} p^{z}_{k^{z}_{\mu}} \bigg) \nonumber \\
     \times\mel{\textbf{e};\textbf{m};  \textbf{c}}{ Z^{\textbf{k}^{z}} X^{\textbf{k}^{x}} X^{\textbf{a}}}{\Psi_0}& \nonumber \\
     \times \mel{\Psi_0}{ (X^{\textbf{a}'})^{\dagger}(X^{\textbf{k}^{x}})^{\dagger}(Z^{\textbf{k}^{z}})^{\dagger} }{ \textbf{e};\textbf{m};  \textbf{b}}. & 
\end{align}
Again these matrix elements are nonzero only if the open Wilson strings in $X^{\textbf{k}^{x}}$ create charges ${e}_v$ at their ends, $(\nabla\cdot \textbf{k}^x)_{v} = e_v \mod N$, and the open 't Hooft strings in $Z^{\textbf{k}^{z}}$ create fluxes ${m}_p$ at their ends, $(\nabla\times \textbf{k}^z)_{p} = m_p \mod N$. Furthermore we must have $\hat{\mathcal{C}}(\textbf{k}^{x})=\textbf{c}-\textbf{a} = \textbf{b}-\textbf{a}'$, otherwise the term vanishes. The new feature relative to the previous calculation is the phase acquired in commuting $X^{\textbf{a}}$ through $Z^{\textbf{k}^{z}}$, which turns $\ket{\textbf{e};\textbf{m};  \textbf{c}}$ into $\ket{\textbf{e};\textbf{m};  \textbf{c}- \textbf{a}}$, and likewise commuting $(X^{\textbf{a}'})^{\dagger}$ through $(Z^{\textbf{k}^{z}})^{\dagger}$, which turns $\ket{\textbf{e};\textbf{m};  \textbf{b}}$ into $\ket{\textbf{e};\textbf{m};  \textbf{b}- \textbf{a}'}\equiv \ket{\textbf{e};\textbf{m};  \textbf{c}- \textbf{a}}$. If $\mathcal{C}(\textbf{k}^{z})=\textbf{d}$, then $Z^{\textbf{k}^{z}} X^{\textbf{a}}= \omega^{\mathbf{d}\cdot\mathbf{a}}X^{\textbf{a}}Z^{\textbf{k}^{z}}$ and $(X^{\textbf{a}'})^{\dagger}(Z^{\textbf{k}^{z}})^{\dagger} = \omega^{-\mathbf{d}\cdot\mathbf{a}'}(Z^{\textbf{k}^{z}})^{\dagger}(X^{\textbf{a}'})^{\dagger}$, so the net phase is $\omega^{\textbf{d}\cdot(\textbf{a}-\textbf{a}')}$ and
 \begin{align}
     \Tr{\mathcal{E}(X_{\textbf{a}}|{\Psi_{0}}\rangle \langle \Psi_{0}| X_{\textbf{a}'}^{\dagger})\rho^{\textbf{b},\textbf{c}}_{\textbf{e};\textbf{m}}}   =& \delta_{\textbf{c}-\textbf{a},\textbf{b}-\textbf{a}'}\nonumber \\
    \times \sum_{\substack{\textbf{k}^{x} \\ (\nabla\cdot \textbf{k}^x)_{v} = e_v \ , \  \hat{\mathcal{C}}(\textbf{k}^{x}) = \textbf{c}-\textbf{a} }} &\prod_{\mu}p^x_{k^{x}_{\mu}} \nonumber \\ 
    \times \sum_{\textbf{d}}\omega^{\textbf{d}(\textbf{a}-\textbf{a}')}\sum_{\substack{\textbf{k}^{z} \\ (\nabla\times \textbf{k}^z)_{p} = m_p \ , \  \mathcal{C}(\textbf{k}^{z}) = \textbf{d} }} &\prod_{\mu}p^z_{k^{z}_{\mu}},  
 \end{align}
with the delta function enforced $\bmod\,N$. The decohered density matrix thus reads
\begin{align}
    \mathcal{E}&(\rho_{QR})  = \sum_{\textbf{a},\textbf{a}'}\sum_{\textbf{g}}\sum_{\substack{\textbf{e};\textbf{m} }} \mathcal{Z}[\textbf{e},\textbf{g}] \nonumber  \\
    &\times \sum_{\textbf{d}}\omega^{\textbf{d}\cdot(\textbf{a}-\textbf{a}')}\mathcal{Z}[\textbf{m},\textbf{d}] \rho^{\textbf{a}+\textbf{g},\textbf{a}'+\textbf{g}}_{\textbf{e};\textbf{m}} \otimes |\textbf{a}\rangle \langle \textbf{a}'|.
\end{align}
Introducing the orthonormal basis of the toric-code-plus-reference Hilbert space,
\begin{align}
    \ket{ \textbf{e};\textbf{m};  \textbf{a},\textbf{b}}_{QR} = \frac{1}{N}\sum_{\textbf{c}} \omega^{\textbf{b}\cdot \textbf{c}}\ket{ \textbf{e};\textbf{m};  \textbf{a}+\textbf{c}}\otimes \ket{\textbf{c}},
\end{align}
the density matrix becomes fully diagonal,
\begin{align}
    \mathcal{E}(\rho_{QR})  = &\sum_{\textbf{a},\textbf{b}}\sum_{\substack{\textbf{e};\textbf{m} }} \mathcal{Z}[\textbf{e},\textbf{a}]\mathcal{Z}[\textbf{m},\textbf{b}]  \nonumber \\
    &| \textbf{e};\textbf{m};  \textbf{a},\textbf{b}\rangle_{QR} \langle  \textbf{e};\textbf{m};  \textbf{a},\textbf{b}|_{QR}.
\end{align}
Once again the spectrum is determined entirely by the partition functions $\mathcal{Z}[\textbf{e},\textbf{a}]$ and $\mathcal{Z}[\textbf{m},\textbf{b}]$.

\section{Deriving \eqnref{eq:QCMI} for the conditional mutual information}\label{app:CMI_appendix}
Let $A$ be a subregion with entangling boundary $\partial A$, which partitions the links into those inside $A$ and those in the complement $\overline{A}$. The boundary $\partial A$ passes through certain vertices, and the Gauss-law operators at these vertices play a special role: such a vertex has links incident from both subregions, so once we restrict to $A$ or $\overline{A}$ alone its Gauss law can appear violated from the perspective of either side. Physically, a nonzero electric charge may then be associated with a boundary vertex when viewed from within $A$ or $\overline{A}$, but it cannot be attributed to either side independently; it is a shared degree of freedom labeling distinct superselection sectors for both subregions. Since gauge-invariant operators supported entirely within $A$ (or $\overline{A}$) cannot modify the charge configuration along $\partial A$, these boundary electric fluxes form the \textit{electric center} of the bipartition algebra~\cite{Donnelly_2012,Donnelly_2014}.
The ground state $\ket{\Psi_0}$ accordingly admits the decomposition
\begin{align}
    \ket{\Psi_0} = \frac{1}{\sqrt{\mathcal{N}_{{\partial A}}}}\sum_{\text{edge modes }\textbf{e}_{\partial A}}\ket{\textbf{e}_{\partial A}}_{A}\ket{ -\textbf{e}_{\partial A}}_{\overline{A}},
\end{align}
where the edge modes $\textbf{e}_{\partial A} = \{{e}_{v \in \partial A}\}$ are the Gauss-law charges on the boundary vertices $v \in \partial A$ and label the superselection sectors. The global Gauss law makes these assignments dependent: for a simply connected region $A$, the total boundary charge must vanish, $\sum_{v \in \partial A}e_{v}=0\mod N$, so the number of independent boundary configurations is $\mathcal{N}_{{\partial A}} = N^{|\partial A|-1}$, where $|\partial A|$ is the number of boundary vertices.
Acting with the $X$-type dephasing channel yields
\begin{align}\label{eq:Full_Decoherence_Channel}
     \mathcal{E}(|\Psi_0\rangle\langle \Psi_0|)  &= \sum_{\{ \textbf{k}^{x}\}} \bigg(\prod_{\mu}p^x_{k^{x}_{\mu}} \bigg)  X^{\textbf{k}^{x}}|\Psi_0\rangle\langle \Psi_0| (X^{\dagger})^{\textbf{k}^{x}}.
\end{align}
Since $X^{\textbf{k}^{x}}$ factorizes across the bipartition, $X^{\textbf{k}^{x}} =X^{\textbf{k}^{x}}_A\otimes X^{\textbf{k}^{x}}_{\overline{A}}$, the state $X^{\textbf{k}^{x}}|\Psi_0\rangle$ decomposes as
\begin{align}
    X^{\textbf{k}^{x}}|\Psi_0\rangle &= \frac{1}{\sqrt{\mathcal{N}_{\partial A}}} \times \nonumber \\
    &\sum_{\text{edge modes }\textbf{e}_{\partial A}} X^{\textbf{k}^{x}}_{A}\ket{\textbf{e}_{\partial A}}_{A}X^{\textbf{k}^{x}}_{\overline{A}}\ket{ -\textbf{e}_{\partial A} }_{\overline{A}},
\end{align}
with $X^{\textbf{k}^{x}}_A = \prod_{\mu\in A}X^{{k}^{x}_{\mu}}_{\mu}$ and $X^{\textbf{k}^{x}}_{\overline{A}} = \prod_{\mu\in \overline{A}}X^{{k}^{x}_{\mu}}_{\mu}$. Tracing out $\overline{A}$ gives
\begin{align}\label{eq:Partial_Trace}
   \Tr_{\overline{A}} X^{\textbf{k}^{x}}|\Psi_0\rangle\langle \Psi_0| (X^{\dagger})^{\textbf{k}^{x}} = \frac{1}{\mathcal{N}_{\partial A}}\sum_{\textbf{e}_{\partial A}}X^{\textbf{k}^{x}}_A |\textbf{e}_{\partial A} \rangle \langle \textbf{e}_{\partial A}|(X^{\textbf{k}^{x}}_A)^{\dagger}.
\end{align}
The reduced density matrix $\rho_A$ commutes with every charge ($A_v$) and flux ($B_p$) operator contained entirely within $A$, as well as with the boundary Gauss-law operators restricted to $A$, whose eigenvalues label the edge modes along $\partial A$. Hence $\mathcal{E}(|\Psi_0\rangle\langle \Psi_0|)$ is diagonal in the basis $\ket{\mathbf{e}_A, \mathbf{e}_{\partial A}}$, where $\mathbf{e}_A$ and $\mathbf{e}_{\partial A}$ denote the bulk and boundary charge configurations in $A$. Repeating the earlier analysis,
\begin{align}\label{eq:Reduced_Density_Matrix}
    \rho_A = \sum_{\textbf{e}_A}\sum_{\textbf{e}_{\partial A}}\frac{\mathcal{Z}[\textbf{e}_A]}{\mathcal{N}_{\partial A}} |\textbf{e}_A,  \textbf{e}_{\partial A}\rangle \langle \textbf{e}_A,  \textbf{e}_{\partial A}|,
\end{align}
where $\mathcal{Z}[\textbf{e}_A] = \sum_{\textbf{e}'_{\partial A} }\mathcal{Z}[\textbf{e}_A, \textbf{e}'_{\partial A} ]$ sums over all boundary charge configurations.
Because \eqnref{eq:Reduced_Density_Matrix} is diagonal, the von Neumann entropy takes the simple form
\begin{align}
    S(A) = (|\partial A|-1)\log N - \sum_{\textbf{e}_{A}}\mathcal{Z}[\textbf{e}_A]\log \mathcal{Z}[\textbf{e}_A],
\end{align}
where the first term is the usual area-law-plus-TEE piece and the second is the additional contribution from coupling to the environment.

Now let $A$ be an annulus with inner boundary $\partial A_{\text{in}}$ and outer boundary $\partial A_{\text{out}}$. The ground state decomposes as
\begin{align}
    \ket{\Psi_0} &= \frac{1}{\sqrt{\mathcal{N}_{\partial A_{\text{in}}}\mathcal{N}_{\partial A_{\text{out}}}}} \times \nonumber \\
    &\sum_{\substack{\textbf{e}_{\partial A_{\text{in}}} , \textbf{e}_{\partial A_{\text{out}}}}}\ket{-\textbf{e}_{\partial A_{\text{in}}}}\ket{ \textbf{e}_{\partial A_{\text{in}}},\textbf{e}_{\partial A_{\text{out}}}}_{A}\ket{ -\textbf{e}_{\partial A_{\text{out}}}},
\end{align}
where $\textbf{e}_{\partial A_{\text{in}}}$ and $\textbf{e}_{\partial A_{\text{out}}}$ are the edge modes on the inner and outer boundaries. A Gauss-law constraint holds on each boundary, $\sum_{v \in \partial A_{\text{in}}}e_v  = \sum_{v \in \partial A_{\text{out}}}e_v = 0\mod N$. Equation~\eqref{eq:Partial_Trace} still applies, but the annulus now has nontrivial homology: $X^{\textbf{k}^{x}}_A$ may, in addition to creating the charge configurations $\mathbf{e}_A$ and $\mathbf{e}_{\partial A}$, also create Wilson lines connecting the inner to the outer boundary. States with the same bulk charges but different net flux threading from inner to outer boundary lie in distinct topological sectors, distinguished by their commutation with the non-contractible 't Hooft loop $Z_{\hat{\gamma}_{\text{ann}}}$ winding around the annulus. Writing $X^{\textbf{k}^{x}}_{A}Z_{\hat{\gamma}_{\text{ann}}} = e^{i2\pi w/N}Z_{\hat{\gamma}_{\text{ann}}}X^{\textbf{k}^{x}}_{A}$, where $w\in\mathbb{Z}_N$ is the net flux of these Wilson lines, we obtain
\begin{align}\label{eq:Reduced_Density_Matrix_Annulus}
    \rho_{A} &= \sum_{\textbf{e}_A, {w}}\sum_{\substack{\textbf{e}_{\partial A_{\text{in}}}  \\ \textbf{e}_{\partial A_{\text{out}}} }} \frac{1}{{\mathcal{N}_{\partial A_{\text{in}}}\mathcal{N}_{\partial A_{\text{out}}}}} \mathcal{Z}[\textbf{e}_A,{w}]\times  \nonumber\\ 
    &|\textbf{e}_{A},\textbf{e}_{\partial A_{\text{in}}}(w),\textbf{e}_{\partial A_{\text{out}}}(w)  \rangle \langle \textbf{e}_{A}, \textbf{e}_{\partial A_{\text{in}}} (w),\textbf{e}_{\partial A_{\text{out}}} (w)|,
\end{align}
where a nonzero $w$ shifts the admissible boundary charges $\textbf{e}_{\partial A_{\text{in}}}(w)$ and $\textbf{e}_{\partial A_{\text{out}}}(w)$ by the flux of the Wilson lines stretching from the inner to the outer boundary. The von Neumann entropy of the annulus is then
\begin{align}
    S(A) = (|\partial A|-2)\log N - \sum_{\textbf{e}_{A},{w}}\mathcal{Z}[\textbf{e}_A, {w}]\log \mathcal{Z}[\textbf{e}_A, {w}].
\end{align}
The crucial new ingredient is the topological number $w$, which has a precise interpretation as the net charge enclosed by the annulus. Applying these formulae to the annular partitions of~\eqref{eq:QCMI_geometry} yields \eqnref{eq:QCMI} for the CMI.
\section{Relative entropy}\label{app:relative_entropy}
The quantum relative entropy is the natural non-commutative generalization of the classical Kullback-Leibler divergence. For density matrices $\rho$ and $\sigma$,
\begin{align}
    D(\rho||\sigma) = \Tr\rho(\ln\rho - \ln \sigma).
\end{align}
It is non-negative, $D(\rho||\sigma) \geq 0$, vanishes iff $\rho = \sigma$, and obeys the data-processing inequality $D(\mathcal{E}(\rho)||\mathcal{E}(\sigma)) \leq D(\rho||\sigma)$ for any channel $\mathcal{E}$ --- the statement that physical operations cannot improve our ability to distinguish states. It is, however, not symmetric.
We are interested in the distinguishability, after decoherence, of two logical states $X^{\bf{a}}\ket{\Psi_0}$ and $X^{\bf{b}}\ket{\Psi_0}$,
\begin{align}\label{eq:Relative_Entropy}
    F^{\bf{a},\bf{b}} = D(\mathcal{E}(\rho_{0}^{\bf{a}})||\mathcal{E}(\rho_{0}^{\bf{b}})),
\end{align}
with $\rho_{0}^{\bf{a}} = X^{\bf{a}}|\Psi_0\rangle \langle \Psi_0|(X^{\bf{a}})^{\dagger}$. Without decoherence the two logical states are orthogonal, so the relative entropy is infinite. If it remains infinite after decoherence (as $L\rightarrow\infty$), the states stay perfectly distinguishable and decoherence can be reversed by the Petz recovery channel; if it drops to a finite value, at least part of the logical information is irretrievably lost and perfect recovery is impossible.
Consider the two logical states $\ket{\Psi_0}$ and $(X^{\bf{a}})^{\dagger}\ket{\Psi_0}$. From the analysis above, the corresponding decohered density matrices are
\begin{align}
    \mathcal{E}((X^{\bf{a}})^{\dagger}|\Psi_0\rangle\langle \Psi_0|X^{\bf{a}}) &= \sum_{\substack{\textbf{e};\textbf{m} \\ \textbf{b}}} \mathcal{Z}[\textbf{e}, \textbf{b}+\textbf{a}] \mathcal{Z}[\textbf{m}] \rho^{\textbf{b},\textbf{b}}_{\textbf{e};\textbf{m}}, \nonumber\\
    \mathcal{E}(|\Psi_0\rangle\langle \Psi_0|)&= \sum_{\substack{\textbf{e};\textbf{m} \\ \textbf{b}}} \mathcal{Z}[\textbf{e}, \textbf{b}] \mathcal{Z}[\textbf{m}] \rho^{\textbf{b},\textbf{b}}_{\textbf{e};\textbf{m}}.
\end{align}
Their relative entropy is
\begin{align}
    F^{\bf{a},\bf{0}} &= \sum_{\substack{\textbf{e}, \textbf{b}}}  \mathcal{Z}[\textbf{e}, \textbf{b}] \log{\frac{ \mathcal{Z}[\textbf{e}, \textbf{b}]}{ \mathcal{Z}[\textbf{e}, \textbf{b}+ \textbf{a}]}}  \nonumber \\
    &= \Big[\, \sum_{\textbf{b}} P[\textbf{b}|\textbf{e}]\,\log\frac{P[\textbf{b}|\textbf{e}]}{P[\textbf{b}+\textbf{a}|\textbf{e}]} \,\Big],
\end{align}
i.e. a disorder-averaged Kullback-Leibler divergence between the conditional winding distributions $P[\textbf{b}|\textbf{e}]$ and $P[\textbf{b}+\textbf{a}|\textbf{e}]$, with $[\cdots]$ the Nishimori-line average. Since $\mathcal{Z}[\textbf{e}, \textbf{b}]$ is the partition function of a disordered spin model with frustrations $\textbf{e}$, $F^{\bf{a},\bf{0}}$ is the disorder-averaged excess free energy of an additional $\mathbf{a}$-twist along the non-contractible cycles of the torus.
\section{Random bond clock models}\label{app:clock_model}
We now present analytic estimates of the coherent information and CMI from the partition function of the random-bond $\mathbb{Z}_{N}$ clock model in its different phases, exploiting the defining property of the Nishimori line that the disorder distribution $P(\bf{J})$ is proportional to the partition function $\mathcal{Z}[\bf{J}]$. For concreteness we keep a single nonzero coupling $\beta_{1} = \beta$, which plays the role of an inverse temperature,
\begin{align}
    H[\{\theta_{i}\};\textbf{k}_\textbf{a}] &= -\beta \sum_{\langle i, j\rangle} \cos\frac{2\pi \big(\theta_{i} - \theta_{j}-k_{{ij}, \textbf{a}}\big)}{N},
\end{align}
on an $L \times L$ lattice with periodic boundary conditions in both directions.
\subsection{Coherent information}
We analyze the coherent information under $X$-type dephasing alone ($Z$-type is identical),
\begin{align}
    I^{(c)}(Q:R;\mathcal{E}) = 2\log N - [ H(\textbf{a}|\textbf{e})],
\end{align}
where $H(\textbf{a}|\textbf{e})$ is the Shannon entropy of $P[\textbf{a}|\textbf{e}]$ and $[\cdots]$ the Nishimori-line disorder average.
\subsubsection{Low-temperature limit}
For $\beta \gg 1$ the $\mathbb{Z}_{N}$ clock model is in the ordered phase, where $\mathcal{Z}[\textbf{e}, \textbf{a}]$ is dominated by minimal-energy saddles. Any bond configuration containing local frustrations $\textbf{e}$ produces a strictly higher-energy saddle, so $\mathcal{Z}$ is exponentially suppressed in $\beta$; since $P(\bf{J})\sim \mathcal{Z}[\bf{J}]$, locally frustrated configurations are exponentially unlikely. This is expected, as the decoherence is weak and charge defects are rare. The same holds for topological frustration: a nonzero $\textbf{a}$ imposes a twist on the torus whose energy cost scales with system size, so that
\begin{align}
    P(\textbf{a}|\textbf{e}) \sim e^{-L \beta\, c_\textbf{a}}, \qquad c_{\textbf{a}} = \sum_{i}\Big(1-\cos\tfrac{2\pi a_{i}}{N}\Big),
\end{align}
which vanishes for $\textbf{a}=0$ and is positive otherwise. The conditional winding entropy therefore scales as
\begin{align}
    [H(\textbf{a}|\textbf{e})]\sim e^{-\alpha L},
\end{align}
giving $I^{(c)}(Q:R;\mathcal{E}) = 2\log N + \mathcal{O}(e^{-L})$. This is the decodable phase.
\subsubsection{High-temperature limit}
For $\beta \ll 1$ the model is in the disordered phase, where frustrations proliferate and energetics are washed out by entropy. Introducing a topological frustration $\textbf{a}$ has little effect, since the associated domain walls are screened by the proliferating vortices. Thus
\begin{align}
    P(\textbf{a}|\textbf{e}) \sim \frac{1}{N^{2}},
\end{align}
giving $I^{(c)}(Q:R;\mathcal{E}) \sim 0$. This is the non-decodable phase.
\subsubsection{QLRO phase}
For $N > 4$ the $\mathbb{Z}_{N}$ clock model has a quasi-long-range-ordered (QLRO) phase, in which the low-energy excitations are gapless spin waves and the infrared theory is Gaussian. We expect the same along the Nishimori line, with bulk disorder merely renormalizing the spin stiffness. Accordingly, we model
\begin{align}
     \mathcal{Z}[\mathbf{e},\mathbf{a}] &= \int\mathcal{D}\vphi\, e^{-H[\vphi,\textbf{J}^{\mathbf{a}}]},\\
    H[\vphi,\textbf{J}] &= \frac{\kappa}{2}\int d^{2}x\,(\nabla \vphi(x) - \textbf{J}^{\mathbf{a}}(x))^{2},
\end{align}
where $\varphi$ is a coarse-grained phase field, effectively non-compact in this phase (vortices are suppressed), $\mathbf{J}^{\mathbf{a}}$ is the coarse-grained disorder field encoding the frustrations $\textbf{e} = \curl \textbf{J}^{\mathbf{a}}$ as well as the topological sector $\textbf{a}$, and $\kappa$ is the effective spin stiffness. Interpreting $\mathbf{E} = -\nabla \varphi$ as an electric field and $\mathbf{J}^{\mathbf{a}}$ as a polarization, the problem maps onto two-dimensional electrostatics: we solve for the electric field in the presence of a background of dipole moments $\mathbf{J}^{\mathbf{a}}(x)$. By the Helmholtz (Hodge) decomposition, any sufficiently smooth vector field on a 2D domain can be written as
\begin{align}
     \textbf{J}(x) =  \textbf{J}_{\text{curl-free}}(x) +  \textbf{J}_{\text{div-free}}(x) + \textbf{J}^{\mathbf{a}}_{\text{harm}}(x),
\end{align}
where
\begin{align}
    \nabla \times\textbf{J}_{\text{curl-free}}(x)=0 &\implies  \textbf{J}_{\text{curl-free}}(x) = \nabla \chi, \\
    \nabla \cdot\textbf{J}_{\text{div-free}}(x)=0  &\implies  \textbf{J}_{\text{div-free}}(x) = \nabla \times\textbf{r}, \\
    \nabla \times\textbf{J}^{\mathbf{a}}_{\text{harm}} &=  \nabla \cdot\textbf{J}^{\mathbf{a}}_{\text{harm}}=0.
\end{align} 
The curl-free part is gauge-dependent and can be absorbed into a redefinition of $\varphi$. The divergence-free component encodes the local bond frustrations, while $\textbf{J}_{\text{harm}}^{\mathbf{a}}$ describes the topological zero modes. In the spin-wave phase vortices are suppressed, so $\nabla\varphi$ is curl-free and cannot screen the divergence-free part of $\mathbf{J}^{\mathbf{a}}$; any such frustration therefore costs energy. Along the Nishimori line frustrated configurations are suppressed and bulk disorder merely renormalizes $\kappa$, just as a polarization field renormalizes the dielectric constant.
The only piece of $\mathbf{J}$ relevant in the spin-wave description is the harmonic mode $\mathbf{J}^{\mathbf{a}}_{\text{harm}}$, which carries the dependence of the free energy on the topological sector $\textbf{a}$,
\begin{align}
    \mathbf{J}^{\textbf{a}}_{\text{harm}} = \frac{\sum_{i}{a_{i}}\hat{e}_{i}}{L},
\end{align}
where $\hat{e}_{i}$ is the unit vector in the $i$th direction and the $a_{i}$ are shifted to the symmetric integer range $\{-\lfloor N/2\rfloor,\ldots,\lfloor N/2\rfloor\}$. Assuming $\varphi$ is periodic on the cylinder, the additional energy is
\begin{align}
    \frac{\kappa}{2}|{\textbf{a}}|^{2} \int_{0}^{L}d^{2}x \,\frac{1}{L^{2}} = \frac{\kappa}{2}|{\textbf{a}}|^{2},
\end{align}
independent of system size. The winding-number probabilities therefore become system-size independent,
\begin{align}
    P(\textbf{a}|\textbf{e}) \sim e^{-\frac{\kappa}{2} |{\textbf{a}}|^{2}},
\end{align}
yielding a fractional coherent information and hence a fractionally decodable phase.
\begin{figure}
    \includegraphics[width=0.85\columnwidth]{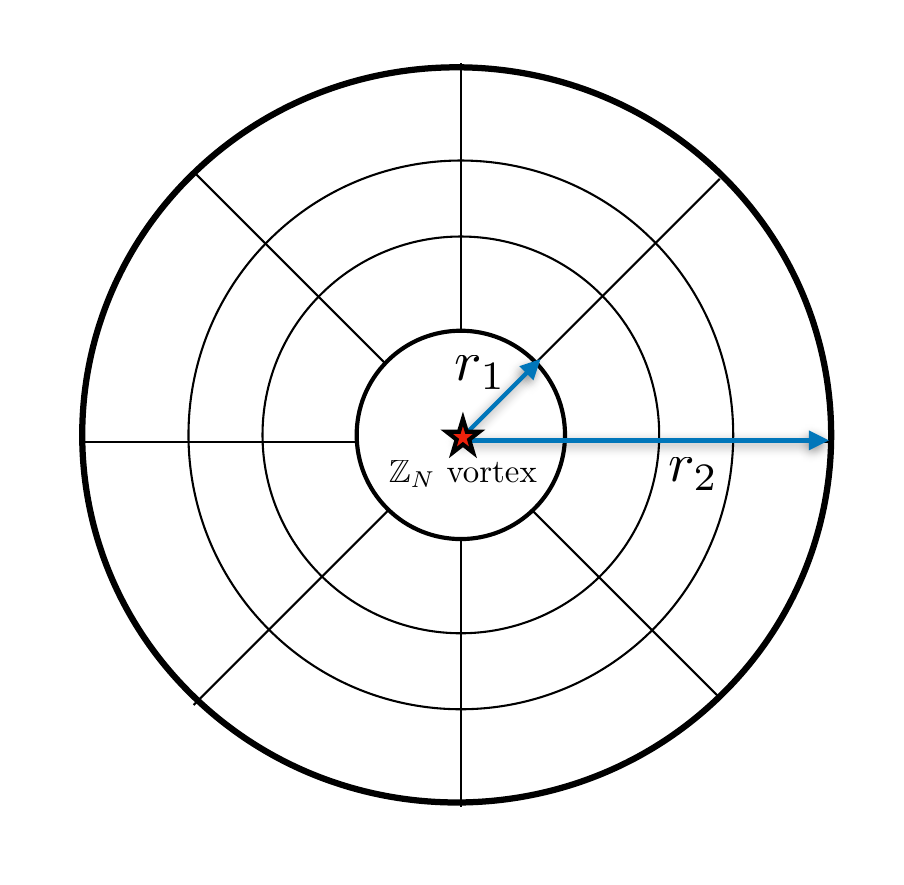}
    \caption{\label{fig:Zn_vortex}  {\bf Partition function on an annulus.} In the QLRO phase, local frustrations remain unimportant but topological modes become important, since their energy no longer scales with system size $L$. On the annulus these modes arise from the nontrivial homology: different winding-number sectors around the annulus correspond to placing different $\mathbb{Z}_{N}$ vortex charges in the hole.} 
\end{figure}
\subsection{CMI}\label{app:CMI_argument}
We now turn to the CMI,
\begin{align}\label{eq:CMI_appendix}
    I(A:C|B) = &- \sum_{\textbf{e}_{AB}}\mathcal{Z}[\textbf{e}_{AB}]\log \mathcal{Z}[\textbf{e}_{AB}] \nonumber \\
    &- \sum_{\textbf{e}_{BC}, E_{A}}\mathcal{Z}[\textbf{e}_{BC}, E_{A}]\log \mathcal{Z}[\textbf{e}_{BC}, E_{A}] \nonumber \\
    &+ \sum_{\textbf{e}_{B},E_{A}}\mathcal{Z}[\textbf{e}_B, E_{A}]\log \mathcal{Z}[\textbf{e}_B, E_{A}] \nonumber \\
    &+ \sum_{\textbf{e}_{ABC}}\mathcal{Z}[\textbf{e}_{ABC}]\log \mathcal{Z}[\textbf{e}_{ABC}],
\end{align}
which are entropies of frustration. In both the ordered and disordered phases the CMI decays exponentially. We now give a qualitative argument, based on the Gaussian spin-wave ansatz, for why it instead saturates in the QLRO phase, implying $\xi_M = \infty$.
Within the spin-wave approximation, bulk disorder is irrelevant and simply renormalizes $\kappa$, exactly as in the ordered phase where frustrations are exponentially suppressed. The first and last terms of~\eqnref{eq:CMI_appendix} therefore do not contribute. The essential difference from the ordered phase is the emergence of the topological zero modes, whose energy no longer scales with system size and which are thus not suppressed as $L\rightarrow \infty$. These modes appear on an annulus through its nontrivial homology: the enclosed charges $E_{A}$ in the $BC$ and $B$ contributions to~\eqnref{eq:CMI_appendix} can be viewed as $\mathbb{Z}_{N}$ vortices placed in the holes of the annulus (\figref{fig:Zn_vortex}). The zero modes they source take the form
\begin{align}
    \mathbf{J}^{\textbf{a}}_{\text{harm}}(r) = \frac{{E_{A}}\hat{e}_{\theta}}{r},
\end{align}
where $\hat{e}_{\theta}$ is the angular unit vector and $E_A$ is shifted to the symmetric range $\{-\lfloor N/2\rfloor,\ldots,\lfloor N/2\rfloor\}$. The partition function for an annulus with inner and outer radii $r_{1}$ and $r_{2}$ and vortex charge $E$ in the hole therefore scales as
\begin{align}
    \mathcal{Z}[E] &\sim \bigg[\frac{r_{1}}{r_{2}}\bigg]^{\pi \kappa {E}^{2}}.
\end{align}
For the annuli $B$ and $BC$ in~\eqnref{eq:CMI_appendix}, with inner/outer radii $(r_A,r_B)$ and $(r_A,r_C)$ respectively,
\begin{align}
    \mathcal{Z}[\textbf{e}_{B}, E_{A}] &\sim \bigg[\frac{r_{A}}{r_{B}}\bigg]^{\pi \kappa {E}_{A}^{2}}, \nonumber \\
    \mathcal{Z}[\textbf{e}_{BC}, E_{A}] &\sim \bigg[\frac{r_{A}}{r_{C}}\bigg]^{\pi \kappa {E}_{A}^{2}}.
\end{align}
Under a proportional rescaling $r_{A,B,C}\rightarrow \lambda r_{A,B,C}$ the zero-mode contributions are unchanged. This scale invariance directly implies that the CMI saturates to a finite, nonzero value, so the Markov length diverges. We therefore conclude that the QLRO phase is an information critical phase.

\end{document}